\documentclass{LMCS}

\usepackage[latin1]{inputenc} 
\usepackage{latexsym, amssymb, amsbsy, stmaryrd,amsmath}
\usepackage{url}
\usepackage{enumerate,hyperref}
\newdimen\proofrulebreadth \proofrulebreadth=.05em
\newdimen\proofdotseparation \proofdotseparation=1.25ex
\newdimen\proofrulebaseline \proofrulebaseline=2ex
\newcount\proofdotnumber \proofdotnumber=3
\let\then\relax
\def\hfi{\hskip0pt plus.0001fil}
\mathchardef\squigto="3A3B
\newif\ifinsideprooftree\insideprooftreefalse
\newif\ifonleftofproofrule\onleftofproofrulefalse
\newif\ifproofdots\proofdotsfalse
\newif\ifdoubleproof\doubleprooffalse
\let\wereinproofbit\relax
\newdimen\shortenproofleft
\newdimen\shortenproofright
\newdimen\proofbelowshift
\newbox\proofabove
\newbox\proofbelow
\newbox\proofrulename
\def\shiftproofbelow{\let\next\relax\afterassignment\setshiftproofbelow\dimen0 }
\def\shiftproofbelowneg{\def\next{\multiply\dimen0 by-1 }%
\afterassignment\setshiftproofbelow\dimen0 }
\def\setshiftproofbelow{\next\proofbelowshift=\dimen0 }
\def\setproofrulebreadth{\proofrulebreadth}

\def\prooftree{
\ifnum  \lastpenalty=1
\then   \unpenalty
\else   \onleftofproofrulefalse
\fi
\ifonleftofproofrule
\else   \ifinsideprooftree
        \then   \hskip.5em plus1fil
        \fi
\fi
\bgroup
\setbox\proofbelow=\hbox{}\setbox\proofrulename=\hbox{}%
\let\justifies\proofover\let\leadsto\proofoverdots\let\Justifies\proofoverdbl
\let\using\proofusing\let\[\prooftree
\ifinsideprooftree\let\]\endprooftree\fi
\proofdotsfalse\doubleprooffalse
\let\thickness\setproofrulebreadth
\let\shiftright\shiftproofbelow \let\shift\shiftproofbelow
\let\shiftleft\shiftproofbelowneg
\let\ifwasinsideprooftree\ifinsideprooftree
\insideprooftreetrue
\setbox\proofabove=\hbox\bgroup$\displaystyle 
\let\wereinproofbit\prooftree
\shortenproofleft=0pt \shortenproofright=0pt \proofbelowshift=0pt
\onleftofproofruletrue\penalty1
}

\def\eproofbit{
\ifx    \wereinproofbit\prooftree
\then   \ifcase \lastpenalty
        \then   \shortenproofright=0pt  
        \or     \unpenalty\hfil         
        \or     \unpenalty\unskip       
        \else   \shortenproofright=0pt  
        \fi
\fi
\global\dimen0=\shortenproofleft
\global\dimen1=\shortenproofright
\global\dimen2=\proofrulebreadth
\global\dimen3=\proofbelowshift
\global\dimen4=\proofdotseparation
$\egroup  
\shortenproofleft=\dimen0
\shortenproofright=\dimen1
\proofrulebreadth=\dimen2
\proofbelowshift=\dimen3
\proofdotseparation=\dimen4
}

\def\proofover{
\eproofbit 
\setbox\proofbelow=\hbox\bgroup 
\let\wereinproofbit\proofover
$\displaystyle
}%
\def\proofoverdbl{
\eproofbit 
\doubleprooftrue
\setbox\proofbelow=\hbox\bgroup 
\let\wereinproofbit\proofoverdbl
$\displaystyle
}%
\def\proofoverdots{
\eproofbit 
\proofdotstrue
\setbox\proofbelow=\hbox\bgroup 
\let\wereinproofbit\proofoverdots
$\displaystyle
}%
\def\proofusing{
\eproofbit 
\setbox\proofrulename=\hbox\bgroup 
\let\wereinproofbit\proofusing
\kern0.3em$
}

\def\endprooftree{
\eproofbit 
  \dimen5 =0pt
\dimen0=\wd\proofabove \advance\dimen0-\shortenproofleft
\advance\dimen0-\shortenproofright
\dimen1=.5\dimen0 \advance\dimen1-.5\wd\proofbelow
\dimen4=\dimen1
\advance\dimen1\proofbelowshift \advance\dimen4-\proofbelowshift
\ifdim  \dimen1<0pt
\then   \advance\shortenproofleft\dimen1
        \advance\dimen0-\dimen1
        \dimen1=0pt
        \ifdim  \shortenproofleft<0pt
        \then   \setbox\proofabove=\hbox{%
                        \kern-\shortenproofleft\unhbox\proofabove}%
                \shortenproofleft=0pt
        \fi
\fi
\ifdim  \dimen4<0pt
\then   \advance\shortenproofright\dimen4
        \advance\dimen0-\dimen4
        \dimen4=0pt
\fi
\ifdim  \shortenproofright<\wd\proofrulename
\then   \shortenproofright=\wd\proofrulename
\fi
\dimen2=\shortenproofleft \advance\dimen2 by\dimen1
\dimen3=\shortenproofright\advance\dimen3 by\dimen4
\ifproofdots
\then
        \dimen6=\shortenproofleft \advance\dimen6 .5\dimen0
        \setbox1=\vbox to\proofdotseparation{\vss\hbox{$\cdot$}\vss}%
        \setbox0=\hbox{%
                \advance\dimen6-.5\wd1
                \kern\dimen6
                $\vcenter to\proofdotnumber\proofdotseparation
                        {\leaders\box1\vfill}$%
                \unhbox\proofrulename}%
\else   \dimen6=\fontdimen22\the\textfont2 
        \dimen7=\dimen6
        \advance\dimen6by.5\proofrulebreadth
        \advance\dimen7by-.5\proofrulebreadth
        \setbox0=\hbox{%
                \kern\shortenproofleft
                \ifdoubleproof
                \then   \hbox to\dimen0{%
                        $\mathsurround0pt\mathord=\mkern-6mu%
                        \cleaders\hbox{$\mkern-2mu=\mkern-2mu$}\hfill
                        \mkern-6mu\mathord=$}%
                \else   \vrule height\dimen6 depth-\dimen7 width\dimen0
                \fi
                \unhbox\proofrulename}%
        \ht0=\dimen6 \dp0=-\dimen7
\fi
\let\doll\relax
\ifwasinsideprooftree
\then   \let\VBOX\vbox
\else   \ifmmode\else$\let\doll=$\fi
        \let\VBOX\vcenter
\fi
\VBOX   {\baselineskip\proofrulebaseline \lineskip.2ex
        \expandafter\lineskiplimit\ifproofdots0ex\else-0.6ex\fi
        \hbox   spread\dimen5   {\hfi\unhbox\proofabove\hfi}%
        \hbox{\box0}%
        \hbox   {\kern\dimen2 \box\proofbelow}}\doll%
\global\dimen2=\dimen2
\global\dimen3=\dimen3
\egroup 
\ifonleftofproofrule
\then   \shortenproofleft=\dimen2
\fi
\shortenproofright=\dimen3
\onleftofproofrulefalse
\ifinsideprooftree
\then   \hskip.5em plus 1fil \penalty2
\fi
}

\usepackage{pst-node, pst-tree}

\usepackage{float}
\floatstyle{boxed}
\newfloat{mafigure}{thp}{pipo}
\floatname{mafigure}{Figure}

\newcommand{\cqfd}{}

\usepackage{ifthen}
\newboolean{extended-abstract}
\setboolean{extended-abstract}{true}

\newcommand{\NE}[1]{\textsf{NE}_{#1}}
\newcommand{\sel}[1]{{\underline {#1}}}
\newcommand{\logeq}{\mbox{$\models\hspace{-0.8mm}\mid$}}
\newcommand{\sem}[1]{[\! [ #1 ]\! ]}
\newcommand{\inference}[2]{\begin{array}{c} #1 \\   \hrulefill \\   #2      \end{array}}
\newcommand{\juxtapose}[2]{\begin{array}{cc} #1 & #2 \end{array}}

\usepackage{ifthen}

\def\frew#1#2#3#4#5#6#7#8{
\setbox0=\hbox{$#6 #7 #1 #8$}%
\setbox1=\hbox{$#6 #7 #2 #8$}%
\ifdim \wd0>\wd1 \rlap{\rlap{\hbox to \wd0{#5}}%
                            {\hbox to\wd0{\hfil\lower #3\box1\relax\hfil}}}{\raise #4\box0}%
\else \rlap{\rlap{\hbox to \wd1{#5}}{\hbox to\wd1{\hfil\raise #4\box0\relax\hfil}}}{\lower #3\box1}%
\fi
}

\def\fstep#1#2#3#4#5{\mathchoice{\frew{#1}{#2}{1.1ex}{1.1ex}{#5}{\scriptstyle}{#3}{#4}}%
                                {\frew{#1}{#2}{0.82ex}{1.1ex}{#5}{\scriptstyle}{#3}{#4}}%
                                {\frew{#1}{#2}{0.51ex}{0.82ex}{#5}{\scriptscriptstyle}{#3}{#4}}%
                                {\frew{#1}{#2}{0.51ex}{0.69ex}{#5}{\scriptscriptstyle}{#3}{#4}}}

\newcommand{\lrstep}[2]{\mathrel{\fstep{#1}{#2}{\;\>}{\>\>\;}{\rightarrowfill}}}

\newcommand{\T}{\mathcal{T}}

\newcommand{\X}{\mathcal{X}}

\newcommand{\R}{\mathcal{R}}

\newcommand{\Part}{\mathcal{P}}
\newcommand{\Q}{\mathcal{Q}}
\newcommand{\final}{\mathsf{f}}
\newcommand{\M}{\mathcal{M}}

\newcommand{\A}{\mathcal{A}}

\newcommand{\vars}{\mathit{vars}}

\newcommand{\Pos}{\mathit{Pos}}

\newcommand{\Id}{\mathit{Id}}

\newcommand{\deter}{\mathit{det}}
\newcommand{\new}{\mathit{new}}

\newcommand{\VTAMC}[2]{$\mathrm{VTAM}^{#1}_{#2}$}
\newcommand{\VTAMS}{\VTAMC{\equiv}{\not\equiv}}
\newcommand{\BTVTAMC}[2]{$\mathrm{BTVTAM}^{#1}_{#2}$}

\newcommand{\PUSH}{\mathsf{PUSH}}
\newcommand{\POP}{\mathsf{POP}}
\newcommand{\INT}{\mathsf{INT}}
\newcommand{\BTINT}{\mathsf{BTINT}}

\def\doi{4 (2:8) 2008}
\lmcsheading%
{\doi}
{1--36}
{}
{}
{Sep.~20, 2007}
{Jun.~18, 2008}
{}

\begin{document}

\title[Visibly Tree Automata with Memory and Constraints]{Visibly Tree
Automata\\ with Memory and Constraints\rsuper *}

\author[H.~Comon-Lundh]{Hubert Comon-Lundh\rsuper a}
\address{{\lsuper a}LSV, CNRS/ENS Cachan}
\email{h.comon-lundh@aist.go.jp}

\author[F.~Jacquemard]{Florent Jacquemard\rsuper b} 
\address{{\lsuper b}INRIA Saclay \& LSV {\scriptsize (CNRS/ENS Cachan)}}
\email{florent.jacquemard@inria.fr}

\author[N.~Perrin]{Nicolas Perrin\rsuper c}
\address{{\lsuper c}ENS Lyon}
\email{nicolas.perrin@ens-lyon.fr}

\titlecomment{{\lsuper *}An extended abstract containing some of the 
results presented in this paper has appeared in the proceeding of FOSSACS'07.}

\begin{abstract}
Tree automata with one memory have been introduced in 2001.
They generalize both pushdown (word) automata and the tree automata
with constraints of equality between brothers of Bogaert and Tison.
Though it has a decidable emptiness problem, 
the main weakness of this model is its lack of good closure properties.

We propose a generalization of the visibly pushdown automata of
Alur and Madhusudan to a family of tree recognizers which carry along 
their (bottom-up) computation an auxiliary unbounded memory with a tree structure 
(instead of a symbol stack).
In other words, these recognizers, called Visibly Tree Automata with Memory
(VTAM) define a subclass of tree automata with one memory 
enjoying Boolean closure properties. 
We show in particular that they can be determinized and 
the problems like emptiness, membership, inclusion and universality are decidable for VTAM.
Moreover, we propose several extensions of VTAM whose
transitions may be constrained by different kinds of  tests between memories
and also constraints a la Bogaert and Tison.
We show that some of these classes of constrained VTAM keep
the good closure  and decidability properties, and we demonstrate their 
expressiveness with relevant examples of tree languages.
\end{abstract}

\maketitle

\section*{Introduction}
The control flow of programs with calls to functions can be abstracted
as pushdown systems.  This allows to reduce some program verification
problems to problems (e.g. model-checking) on pushdown automata.  When
it comes to functional languages with \emph{continuation passing
style}, the stack must contain information on continuations and has
the structure of a dag (for jumps).  Similarly, in the context of
asynchronous concurrent programming languages, for two concurrent
threads the ordering of return is not determined (synchronized) and
these threads can not be stacked.  In these cases, the control flow is
better modeled as a tree structure rather than a stack.
That is why we are interested in tree automata with one memory, which
generalize the pushdown (tree) automata, replacing the a stack with a  tree.
Here, a ``memory'' has to be understood as a storage device, whose structure
is a tree. For instance, two memories would correspond to two storage devices
whose access would be independent. 

The \emph{tree automata with one memory} introduced in~\cite{ComonCortier05tcs}
compute bottom-up on a tree, with an auxiliary memory carrying a tree,
as in former works such as \cite{Guessarian83}.  
Along a computation, at any node of the tree, the memory is
updated incrementally from the memory reached at the sons of the node.
This update may consist in building a new tree from the memories at
the sons (this generalizes a push) or retrieving a subtree of one of
the memories at the sons (this generalizes a pop). In addition, such
automata may perform equality tests: a transition may be constrained to
be performed, only when the memories reached at some of the sons are
identical. In this way, tree automata with one memory also generalize
certain cases 
of tree automata with equality and disequality tests between brothers~\cite{BogaertTison92}. 

Automata with one memory have been introduced in the context of the
verification of security protocols, where the messages exchanged are
represented as trees.  In the context of (functional or concurrent) 
programs,  the creation of a thread, or a \textsf{callcc},
 corresponds to a push, the 
termination of a thread or a \textsf{callcc} corresponds to a pop.
The emptiness problem for such automata is in EXPTIME
(note that for the extension with a second memory
 the emptiness problem becomes undecidable).
However, the class of tree languages defined by such automata is neither closed by
intersection nor by complement. This is not surprising as they are 
strictly more general than context free languages. 

On the other hand, Alur and Madhusudan have introduced the notion of
visibility for pushdown automata~\cite{AlurMadhusudan04VPL}, which is
a relevant restriction in the context of control flow analysis.  With
this restriction, determinization is possible and actually the class
of languages is closed under Boolean operations.

\bigskip
In this paper, we propose the new formalism of Visibly Tree Automata
with Memory (VTAM). On one hand, it extends visibly pushdown
languages to the recognition of trees, 
and with a tree structure instead of a stack,
following former approaches \cite{Guessarian83,SchimpfGallier85,CoquideDauchetGilleronVagvolgyi94}.
On the other hand, VTAM restrict tree automata with one memory, imposing a
visibility condition on the transitions: each symbol is assigned a
given type of action. When reading a symbol, the automaton can only
perform the assigned type of action: push or pop.

We first show in Section~\ref{sec:vtam} that VTAM can be
determinized, using a proof similar to the proof of
\cite{AlurMadhusudan04VPL}, and do have the good closure
properties. The main difficulty here is to understand what is a good
notion of visibility for trees, with memories instead of stacks.
We also show that the problems of membership and emptiness are
decidable in deterministic polynomial time for VTAM.

In a second part of the paper (Section~\ref{sec:vtamc}), we
extended VTAM with constraints. Our constraints here are recognizable
relations; a transition can be fired only if the memory contents of
the sons of the current node satisfy such a relation.  
We give then a general theorem, expressing conditions on the relations, which ensure
the decidability of emptiness.  Such conditions are shown to be
necessary on one hand, and, on the other hand, we prove that they are
satisfied by some examples, including syntactic equality and disequality tests 
and structural equality and disequality tests. 
The case of VTAM with structural equality and disequality tests 
(this class is denoted \VTAMS) is particularly interesting, 
since the determinization and closure properties of Section~\ref{sec:vtam} 
carry over this generalization, which we show in Section~\ref{sec:constraint-structural}.
The automata of \VTAMS{} also enjoy a good expressive power, 
as we show in Section~\ref{sec:examples} by presenting some non-trivial examples of
languages in this class: well-balanced binary trees,  red-black trees, powerlists...

As an intermediate result, we show that, in case of
equality tests or structural equality tests, the language of memories
that can be reached in a given state is always a regular language.
This is a generalization of the well-known result that the set of
stack contents in a pushdown automaton is always regular. To prove
this, we observe that the memories contents are recognized by a
two-way alternating tree automaton with constraints. Then we show,
using a saturation strategy, that two-way alternating tree automata
with (structural) equality constraints are not more expressive than
standard tree automata. 

Finally, in Section~\ref{sec:BT} we propose a class of visibly tree automata,
which combines the structural constraints of \VTAMS{}, testing memory contents,
with Bogaert-Tison constraints of~\cite{BogaertTison92} 
(equality and disequality tests between brothers subterms)
which operate on the term in input.
We show that the tree automata of this class can be determinized,
are closed under Boolean operations and have a decidable emptiness problem.

\subsection*{Related Work} 
Generalizations of pushdown automata to trees 
(both for input  and stack) are proposed in 
\cite{Guessarian83,SchimpfGallier85,CoquideDauchetGilleronVagvolgyi94}.
Our contributions are the generalization of the visibility condition of~\cite{AlurMadhusudan04VPL}
to such tree automata -- our VTAM (without constraints) strictly generalize
the VP Languages of~\cite{AlurMadhusudan04VPL},
and the addition of constraints on the stack contents.
The visibly tree automata of~\cite{AlurChaudhuriMadhusudan06VPTL}
use a word stack which is less general than a tree structured memory
but the comparison with VTAM is not easy as 
they are alternating and compute top-down on infinite trees.

Independently, Chabin and Rety have proposed~\cite{ChabinRety07} a 
formalism combining pushdown tree automata of~\cite{Guessarian83}
with the concept of visibly pushdown languages.
Their automata recognize finite trees using a word stack.
They have a decidable  emptiness problem and
the corresponding tree languages (Visibly Pushdown Tree Languages, VPTL) 
are closed under Boolean operations.
Following remarks of one of  these two authors, it appeared that 
VTAM and VPTL are incomparable, see Section~\ref{sec:expressiveness}.



\section{Preliminaries}
\subsection{Term algebra}
A \emph{signature} $\Sigma$ is a finite set of function symbols with arity, denoted by $f$, $g$\ldots 
We write $\Sigma_n$ the subset of function symbols of $\Sigma$ of arity $n$.
Given an infinite set  $\X$ of variables, the set of terms built over $\Sigma$ and $\X$ 
is denoted $\T(\Sigma, \X)$, and the subset of ground terms is denoted $\T(\Sigma)$.
%
The set of variables occurring in a term $t\in \T(\Sigma, \X)$ is denoted $\vars(t)$.
A \emph{substitution} $\sigma$ is a 
mapping from $\X$ to $\T(\Sigma, \X)$ such that $\{x|\sigma(x)~\not=x \}$,
the \emph{support} of $\sigma$, is finite.  The application of a
substitution $\sigma$ to a term $t$ is written $t\sigma$.
It is the homomorphic extension of $\sigma$ to $\T(\Sigma, \X)$.
The \emph{positions} $\Pos(t)$ in a term $t$ are sequences of
positive integers ($\Lambda$, the empty sequence, is the root position).  
A subterm of $t$ at position $p$ is written $t|_p$, and
the replacement in $t$ of the subterm at position $p$ by $u$ denoted $t[u]_p$.

\subsection{Rewriting} 
We assume standard definitions and notations for term rewriting~\cite{DershowitzJouannaud90}.
A \emph{term rewriting system} (TRS) over a signature $\Sigma$ is a finite set of rewrite rules
$\ell \to r$, where $\ell \in \T(\Sigma, \X)$ and $r\in \T(\Sigma,\vars(\ell))$. 
A term $t \in \T(\Sigma, \X)$ rewrites to $s$ by a TRS  $\R$  (denoted $t \to_\R s$)
if there is a rewrite rule $\ell \to r \in \R$, a position
$p$ of $t$ and a substitution $\sigma $ such that $t|_p =\ell\sigma$ and 
$s= t[r\sigma]_p$. 
The transitive and reflexive closure of $\to_\R$ is denoted $\lrstep{*}{\R}$.

\subsection{Tree Automata}
Following definitions and notation of~\cite{tata},
we consider tree automata which compute bottom-up 
(from leaves to root) on (finite) ground terms in $\T(\Sigma)$.
%
At each stage of computation on a tree $t$, a tree automaton 
reads the function symbol $f$ at the current position $p$ in $t$ and
updates its current state, according to $f$ and 
to the respective states reached at the 
positions immediately under $p$ in $t$.
Formally, a bottom-up \emph{tree automaton} (TA) $\A$ 
on a signature $\Sigma$ is a tuple $(Q, Q_\final, \Delta)$ where
$\Sigma$ is the computation signature, 
$Q$ is a finite set of nullary state symbols, disjoint from $\Sigma$,
$Q_\final \subseteq Q$ is the subset of final states and 
$\Delta$ is a set of rewrite rules of the form:
$f(q_1, \ldots, q_n) \to q$,
where $f \in \Sigma$ 
and $q_1,\ldots, q_n \in Q$.
%
A term $t$ is \emph{accepted} (we may also write \emph{recognized}) 
by $\A$ in state $q$ iff $t \lrstep{*}{\Delta} q$,
and the \emph{language} $L(\A, q)$ of $\A$ in state $q$ is the set of 
ground terms accepted in $q$.
The language $L(\A)$ of  $\A$ is $\bigcup_{q \in Q_\final} L(\A, q)$
and a set of ground terms is called \emph{regular} if it is the language of a TA.

\section{Visibly Tree Automata with Memory} \label{sec:vtam}
We propose in this section a subclass of the tree automata 
with one memory~\cite{ComonCortier05tcs}
which is stable under Boolean operations and has decidable emptiness and membership problems.

\subsection{Definition of VTAM} \label{sec:vtam-definition}
Tree automata have been 
extended~\cite{Guessarian83,SchimpfGallier85,CoquideDauchetGilleronVagvolgyi94,ComonCortier05tcs} 
to carry an unbounded information
along the states in computations.
In~\cite{ComonCortier05tcs},
this information is stored in a tree structure 
and is called \emph{memory}.
We keep this terminology here, and call our recognizers
\emph{tree automata with memory} (TAM).
For consistency with the above formalisms,
the memory contents will be ground terms over a 
\emph{memory signature}~$\Gamma$.

Like for TA we consider bottom-up computations of TAM  in trees;
at each stage of computation on a tree $t$, 
a TAM, like a TA,  reads the function symbol at the current position $p$ in $t$ and
updates its current state, according to the states reached immediately under $p$.
Moreover, a configuration of TAM contains not only a state
but also a memory, which is a tree. 
The current memory is updated according to the respective contents 
of memories reached in the nodes immediately under $p$ in $t$.

As above, we use term rewrite systems in order to define the transitions allowed in a TAM.
For this purpose, we add an argument to state symbols, which will
contain the memory. Hence, a configuration of TAM
in state $q$ and whose memory content is the ground term $m \in \T(\Gamma)$,
is represented by the term $q(m)$.
We propose below a very general definition of TAM. It is similar
to the  one of~\cite{ComonCortier05tcs}, except that we have here
general patterns $m_1,\ldots,m_n,m$, while these patterns are
restricted in \cite{ComonCortier05tcs}, for instance avoiding memory
duplications. Since we aim at providing closure and decision properties,
we will also impose (other) restrictions later on.
\begin{defi} \label{def:tam}
A bottom-up \emph{tree automaton with memory} (TAM) 
on a signature $\Sigma$ 
is a tuple $(\Gamma, Q, Q_\final, \Delta)$ where
$\Gamma$ is a memory signature, 
$Q$ is a finite set of unary state symbols, disjoint from $\Sigma \cup \Gamma$,
$Q_\final \subseteq Q$ is the subset of final states and 
$\Delta$ is a set of rewrite rules of the form
$f\bigl(q_1(m_1), \ldots, q_n(m_n)\bigr)  \to q(m)$
where 
 $f \in \Sigma_n$, $q_1,\ldots,q_n, q \in Q$ and $m_1,\ldots, m_n, m \in \T(\Gamma, \X)$.
\end{defi}
%
The rules of $\Delta$ are also called \emph{transition rules}.
A term $t$ is \emph{accepted} by $\A$ in state $q \in Q$ and with memory
$m \in \T(\Gamma)$ iff $t \lrstep{*}{\Delta} q(m)$, and the 
\emph{language} $L(\A, q)$ and \emph{memory language} $M(\A, q)$
of $\A$ in state $q$ are respectively defined by:
\[
\begin{array}{rclcll}
L(\A, q) & = & \bigl\{ t & \bigm| & \exists m \in \T(\Gamma), & t \lrstep{*}{\Delta} q(m) \bigr\}\\[1mm]
M(\A, q) & = & \bigl\{ m & \bigm| & \exists t \in \T(\Sigma),   & t \lrstep{*}{\Delta} q(m) \bigr\}.
\end{array}
\]
The language of $\A$ is the union of languages of $\A$ in its final states, denoted:
\( L(\A) = \bigcup_{q \in Q_\final} L(\A, q). \)



\subsection*{Visibility Condition}
The above formalism is of course far too expressive.
As there are no restrictions on the operation performed on 
memory by the rewrite rules, one can easily encode a Turing machine as a TAM.
We shall now define a decidable restriction called \emph{visibly tree automata with memory} (VTAM).

First, we consider only three main families 
(later divided into the subcategories defined in Figure~\ref{fig:categories}) 
of operations on memory.
We assume below a computation step at some position $p$ of a term, 
where memories $m_1,\ldots, m_n$ have been reached at the positions
immediately below $p$:
\begin{enumerate}[\ ]
\item {\hskip-8 pt$\PUSH$:}
the new current memory $m$ is built with a symbol $h \in \Gamma_n$ \emph{pushed}
on the top of memories $m_1,\ldots,m_n$:
$f \bigl(q_1(m_1),  \ldots, q_n(m_n)\bigr)  \to q\bigl(h(m_1, \ldots, m_n)\bigr)$.
%
According to the terminology of~\cite{AlurMadhusudan04VPL}, 
this corresponds to a \emph{call} move in a program represented by an automaton.
\item {\hskip-8 pt$\POP$:}
the new current memory is a subterm of one of 
the  memories reached so far:
$f\bigl( \ldots, q_i(h(m'_1,\ldots, m'_k)),  \ldots\bigr) \to q(m'_j)$.
The top symbol $h$ of $m_i$ is also read.
%
This corresponds to a function's \emph{return} in a program.

We have here to split $\POP$ operations into four categories, 
depending on whether we
pop on the memory at the left son or on the memory at the right son 
and on whether we get the left son of that memory or its right son.

\item {\hskip-8 pt$\INT$ {\bf (internal)}:}
the new current memory is  one of the memories reached: 
\[f \bigl(q_1(m_1),  \ldots, q_n(m_n)\bigr)  \to q(m_i)\]
%
This corresponds to an internal operation (neither call nor return) 
in a function of a program.

Again, we need to split $\INT$ operations into three categories: 
one for constant symbols and 
two rules for binary symbols, depending on which of the two sons memories
we keep.
\end{enumerate}

Next, we adhere to the \emph{visibility} condition of~\cite{AlurMadhusudan04VPL}.
The idea behind this restriction,
which was already in~\cite{JensenMetayerThorn99},
is that the symbol read by an automaton
(in a term in our case  and~\cite{AlurChaudhuriMadhusudan06VPTL},
in a word in the case of~\cite{AlurMadhusudan04VPL}) 
corresponds to an instruction of  a program, and hence belongs to
one of the three above families (call, return or internal). 
Indeed, the effect of the execution of a given instruction on the current program state
(a stack for~\cite{AlurMadhusudan04VPL} or a tree in our case)
will always be in the same family.
In other words, in this context, the family of the memory operations performed 
by a transition is completely determined by the function symbol read.

Let us assume from now on for the sake of simplicity the following restriction
on the arity of symbols:
\begin{quote}
All the symbols of $\Sigma$ and $\Gamma$ have either arity 0 or 2.
\end{quote}
This is not a real restriction, and the results of this paper can be extended
straightforwardly to the case of function symbols with other arities. 
The signature $\Sigma$ is partitioned in eight subsets:
\label{def:VTAM-partition}
	\[  \Sigma = 
	     \Sigma_{\PUSH} 
		\uplus \Sigma_{\POP_{11}}  \uplus \Sigma_{\POP_{12}} \uplus \Sigma_{\POP_{21}} \uplus \Sigma_{\POP_{22}}
		\uplus \Sigma_{\INT_{0}} \uplus \Sigma_{\INT_{1}} \uplus \Sigma_{\INT_{2}} \]
The eight corresponding categories of transitions
(transitions of the same category perform the same kind of operation on the memory)
are defined formally in Figure~\ref{fig:categories}.
In this figure, one constant symbol has a particular role:
\begin{quote}
$\bot$ is a special constant symbol in $\Gamma$, used to represent an empty memory. 
\end{quote}
Note that there are three categories for $\INT$, 
$\INT_0$ is for constant symbols and
$\INT_1$, $\INT_2$ are for binary symbols and differ
according to the memory which is kept.
Similarly, there are four variants of $\POP$ transitions, 
$\POP_{11},\ldots,\POP_{22}$.
Moreover, each $\POP$ rule has a variant, which reads an empty memory (\emph{i.e.} the symbol $\bot$).
%
\begin{figure}
\[
\begin{array}{lcllclcl}
\PUSH_{\phantom{0}} & & a  & & \to & q(c)  & \quad & a \in \Sigma_\PUSH\\
\PUSH_{\phantom{0}} & & f \bigl(q_1(y_1), & q_2(y_2)\bigr)  & \to & q\bigl(h(y_1, y_2)\bigr) & & f \in \Sigma_\PUSH\\
\POP_{11} & & f\bigl(q_1(h(y_{11}, y_{12})), & q_2(y_2)\bigr)  & \to & q(y_{11}) & & f \in \Sigma_{\POP_{11}}\\
                    & & f\bigl(q_1(\bot), & q_2(y_2)\bigr)  & \to & q(\bot)\\
\POP_{12} & & f\bigl(q_1(h(y_{11}, y_{12})), & q_2(y_2)\bigr)  & \to & q(y_{12}) & & f\in \Sigma_{\POP_{12}}\\
                    & & f\bigl(q_1(\bot ), & q_2(y_2)\bigr)  & \to & q(\bot)\\
\POP_{21} & & f\bigl(q_1(y_1), & q_2(h(y_{21}, y_{22}))\bigr)  & \to & q(y_{21}) & & f \in \Sigma_{\POP_{21}}\\
                    & & f\bigl(q_1(y_1), & q_2(\bot) \bigr)  & \to & q(\bot)\\
\POP_{22} & & f\bigl(q_1(y_1), & q_2(h(y_{21}, y_{22}))\bigr)  & \to & q(y_{22}) & & f \in \Sigma_{\POP_{22}}\\
                    & & f\bigl(q_1(y_1), & q_2(\bot)\bigr)  & \to & q(\bot)\\
\INT_0 & & a  &   & \to & q(\bot)  & & a \in \Sigma_{\INT_0}\\
\INT_1 & & f\bigl(q_1(y_1), & q_2(y_2)\bigr)  & \to & q(y_1) & & f \in \Sigma_{\INT_1}\\
\INT_2 & & f\bigl(q_1(y_1), & q_2(y_2)\bigr)  & \to & q(y_2) & & f\in \Sigma_{\INT_2}\\
\end{array}
\]
\begin{center}
where $q_1, q_2, q \in Q$, $y_1, y_2$ are distinct variables of $\X$,
$c \in \Gamma_2$,
$h \in \Gamma_2$. 
\end{center}
\caption{VTAM transition categories.}
\label{fig:categories}
\end{figure}

\begin{defi} \label{def:vtam}
\sloppypar A \emph{visibly tree automaton with memory} (or VTAM for short) 
on $\Sigma$ 
is a TAM $(\Gamma, Q, Q_\final, \Delta)$ such that 
every rule of $\Delta$ belongs to one of the 
above categories 
$\PUSH$, 
$\POP_{11}$, 
$\POP_{12}$, 
$\POP_{21}$, 
$\POP_{22}$, 
$\INT_0$,
$\INT_1$,
$\INT_2$.
\end{defi}


\subsection{Expressiveness, Comparison}  \label{sec:expressiveness}
Standard bottom-up tree automata are particular cases of VTAM
(simply assume all the symbols of the signature in $\INT_0$ or $\INT_1$).

Now, let us try to explain more precisely the relation with the
visibly pushdown languages of \cite{AlurMadhusudan04VPL}, when
considering finite word languages.

If the stack is empty in any accepting configuration of some finite word
pushdown automaton $\A$, then it is easy to compute a pushdown automaton 
$\widetilde{\A}$,
which accepts the reverses (mirror images) 
of the words accepted by $\A$. 
Moreover, if $\A$ is a visibly pushdown automaton, 
then $\widetilde{\A}$ is also a visibly pushdown automaton: 
it suffices to exchange the push and pop symbols. 

For pushdown word languages, there is a well-known lemma showing that
the recognition by final state is equivalent to the recognition by
empty stack. This equivalence however requires $\epsilon$-transitions 
to empty the stack when a final state is reached. 
There are however
no $\epsilon$-transitions in visibly pushdown automata. 
So, if we consider for instance the language of words 
$w\in \{a,b\}^*$ 
such that any prefix of $w$ contains more $a$ than $b$'s,
it is recognized by a visibly pushdown automaton. While, if we
consider the mirror image (all suffixes contain more $a$'s than $b$'s),
it is not recognized by a visibly pushdown automaton.

In conclusion, as long as visibility is relevant, the way the automaton
is moving is also relevant. 
This applies of course to trees as well: there is a difference between
top-down and bottom-up recognition. 

Now, if we encode a word as a tree on a unary alphabet, starting from right
to left, VTAM generalize visibly pushdown automata: 
moving bottom-up in the tree corresponds
to moving left-right in the word.

VPTA transitions and VPTL are defined in~\cite{ChabinRety07}
in the same formalism (rewrite rules) as in Figure~\ref{fig:categories},
except that the rules are oriented in the other direction 
(top-down computations) and the memory contains a word,  i.e.  terms built with 
unary function symbols and one constant (empty stack).

As sketched above, since the automata of \cite{ChabinRety07} work top-down,
a language can be recognized by a VTAM (which works bottom-up) and not
by a VPTL. As a typical example, consider the trees containing only unary
symbols $a,b$ and a constant $0$ and such that all subterms contain more
$a$'s than $b$'s. 

But the converse is also true: there are similarly languages that are
recognized by VPTA and not by VTAM (and there, constraints cannot help!)

Now, if we consider a slight modification of VPTA, in which the automata
work bottom-up (simply change the direction of transition rules),
it is not clear that good properties (closure and decision) are preserved
since, now, we get equality tests between memory contents, increasing
the original expressive power; when going top-down we always duplicate
the memory content and send one copy to each son, while going bottom-up
we may have different memory contents at two brother
positions. 

\subsection{Determinism}    \label{sec:vtam-determinism}
A VTAM $\A$ is said \emph{complete} if every term of $\T(\Sigma)$ 
belongs to $L(\A, q)$
for at least one state $q \in Q$.
Every VTAM can be completed (with a polynomial overhead) by the addition
of a trash state. 
Hence, we shall consider from now on only complete VTAM.

\bigskip

\noindent A VTAM $\A = (\Gamma, Q, Q_\final, \Delta)$ 
is said \emph{deterministic} iff:
\begin{enumerate}[$\bullet$]
\item for all $a \in \Sigma_{\INT_0}$ there is at most one rule in $\Delta$ with left-member $a$,
\item for all $f \in \Sigma_{\PUSH} \cup \Sigma_{\INT_1} \cup \Sigma_{\INT_2}$, for all $q_1, q_2 \in Q$, 
	there is at most one rule in $\Delta$ with left-member $f\bigl(q_1(y_1), q_2(y_2)\bigr)$,
\item for all $f \in \Sigma_{\POP_{11}} \cup \Sigma_{\POP_{12}}$ 
	(respectively $\Sigma_{\POP_{21}} \cup \Sigma_{\POP_{22}}$),
	 for all $q_1, q_2 \in Q$ and all $h \in \Gamma$, 
	there is at most one rule in $\Delta$ with left-member 
    $f\bigl(q_1(h(y_{11}, y_{12})), q_2(y_2)\bigr)$  (respectively $f\bigl(q_1(y_1), q_2(h(y_{21}, y_{22}))\bigr)$).
\end{enumerate}

\begin{thm} \label{th:vtam-determinization}
For every VTAM $\A = (\Gamma, Q, Q_\final, \Delta)$ there exists a deterministic 
VTAM $\A^\deter = (\Gamma^\deter, Q^\deter, Q_\final^\deter, \Delta^\deter)$ such that $L(\A) = L(\A^\deter)$, 
where $| Q^\deter |$ and $| \Gamma^\deter |$ both are $O\bigl(2^{|Q|^2}\bigr)$.
\end{thm}
\begin{proof}  
We follow the technique of~\cite{AlurMadhusudan04VPL} 
for the determinization of visibly pushdown automata:
we do a subset construction and postpone the application (to the memory) 
of $\PUSH$ rules,
until a matching $\POP$ is met. 
%
The construction of~\cite{AlurMadhusudan04VPL} 
is extended in order to handle the branching structure of the term read and of the memory.

With the visibility condition, for each symbol read, only one kind of memory operation is possible.
This permits a uniform construction of the rules of $\A^\deter$ for each symbol of $\Sigma$.
As we shall see below, 
$\A^\deter$ does not need to keep track of the contents of memory (of $\A$) during its computation, 
it only needs to memorize information on the reachability of  states of $\A$,
following the path (in the term read) from the position of the $\PUSH$ symbol which has 
pushed the top symbol of the current memory (let us call it the \emph{last-memory-push-position})
to the current position in the term.
%
We let :
\[
Q^\deter := \{ 0, 1\} \times \Part(Q) \times \Part(Q^2)
\]
$Q_\final^\deter$ is the subset of states whose second component contains a final state of $Q_\final$.
The first component is a flag indicating whether the memory is currently empty (value 0) or not (value 1).
The second component is the subset of states of $Q$ that $\A$ can reach at current position, and
the third component is a binary relation on $Q$ which contains $(q, q')$ iff
starting from a state $q$ and memory $m$ at the {last-memory-push-position}, 
$\A$ can reach the current position in state $q'$, and with the same memory $m$.
%
\noindent We consider memory symbols made of pairs of states and $\PUSH$ symbols:
\[
\Gamma^\deter := \bigl(Q^\deter\bigr)^2 \times ( \Sigma_{\PUSH} )
\]
The components of a symbol $p \in \Gamma^\deter$ refer
to the transition who pushed $p$:
the first and second components of $p$ are
respectively the left and right initial states of the transition
and the third component is the symbol read.

The transition rules of $\Delta^\deter$ are given below, according to the
symbol read.

\paragraph{$\INT$.} For every $i$ and for every $f \in \Sigma_{\INT_i}$, 
we have the following rules in $\Delta^\deter$:
\[ 
f\bigl( \langle b_1, R_1, S_1\rangle(y_1), \langle b_2, R_2, S_2\rangle(y_2) \bigr)
\to \langle b_1, R, S\rangle(y_1)
\]
where 
\( R := \bigl\{ q \bigm| 
   \exists q_1 \in R_1, q_2 \in R_2, f\bigl(q_1(y_1), q_2(y_2)\bigr) \to q(y_1) \in \Delta \bigr\}\),
and $S$ is the update of $S_1$ according to the $\INT_1$-transitions of $\Delta$, 
when $b_1 = 1$ (the case $b_1 = 0$ is similar):
\[ S := \bigl\{ (q, q') \bigm| \exists q_1 \in Q, q_2 \in R_2, (q, q_1) \in S_1
\mbox{~and~}  f\bigl(q_1(y_1), q_2(y_2)\bigr) \to q'(y_1) \in \Delta  \bigr\}. \]
The case $f \in \Sigma_{\INT_2}$ is similar.

\paragraph{$\PUSH$.} For every $f \in \Sigma_{\PUSH}$, we have the following rules in $\Delta^\deter$:
\[ 
f\bigl( \langle b_1, R_1, S_1\rangle(y_1), \langle b_2, R_2, S_2\rangle(y_2) \bigr)
\to \langle 1, R, \Id_Q \rangle(p(y_1, y_2))
\]
where \( R := \bigl\{ q  \bigm|  \exists q_1 \in R_1, q_2 \in R_2, h \in \Gamma,
         f\bigl(q_1(y_1), q_2(y_2)\bigr) \to q\bigl(h(y_1, y_2)\bigr) \in \Delta \bigr\}\),
$\Id_Q := \bigl\{ (q, q) \bigm| q \in Q \bigr\}$ is used to initialize the memorization of state
reachability from the position of the symbol $f$, 
and $p := \bigl\langle \langle b_1, R_1, S_1\rangle, \langle b_2, R_2, S_2\rangle, f \bigr\rangle$.
Note that the two states reached just below the position of application of this rule are pushed on the
top of the memory. They will be used later in order to update $R$ and $S$ when a matching $\POP$ symbol is read.

\paragraph{$\POP$.} For every $f \in \Sigma_{\POP_{11}}$, we have the following rules in $\Delta^\deter$:
\[ 
f\bigl( \langle b_1, R_1, S_1\rangle(H(y_{11}, y_{12})),  \langle b_2, R_2, S_2\rangle(y_2) \bigr)
\to \langle b, R, S \rangle(y_{11})
\]
where $H = \langle Q_1, Q_2, g \rangle$, with 
$Q_1 = \langle b'_{1}, R'_{1}, S'_{1} \rangle \in Q^\deter$, 
$Q_2 = \langle b'_{2}, R'_{2}, S'_{2} \rangle \in Q^\deter$.
\[
\begin{array}{rcl}
b & = & \quad b'_1\\
R & = & \left\{ \phantom{(}q\phantom{, q')}  \left| 
\begin{array}{l}
\exists q'_1 \in R'_1, q'_2 \in R'_2, (q_0, q_1) \in S_1, q_2 \in R_2,  h\in \Gamma,
 g \bigl(q'_1(y_1),  q'_2(y_2)\bigr)  \to \\
 \quad q_0\bigl(h(y_1, y_2)\bigr) \in \Delta,
f\bigl(q_1(h(y_{11}, y_{12})), q_2(y_2)\bigr)  \to q(y_{11}) \in \Delta
\end{array}
\right. \right\} 
\\[3mm]
S & = & \left\{ (q, q')  \left|
\begin{array}{l}
\exists q'_1 \in S'_1(q), q'_2 \in R'_2, (q_0, q_1) \in S_1, q_2 \in R_2,  h\in \Gamma,
 g \bigl(q'_1(y_1),  q'_2(y_2)\bigr)  \\
 \to q_0\bigl(h(y_1, y_2)\bigr) \in \Delta,
f\bigl(q_1(h(y_{11}, y_{12})), q_2(y_2)\bigr)  \to q'(y_{11}) \in \Delta
\end{array}
\right. \right\}
\end{array}
\]
When a $\POP$ symbol is read, the top symbol of the memory, which is popped,
contains the states reached just before the application of the matching $\PUSH$.
We use this information in order to update 
$\langle b_1, R_1, S_1\rangle$ and $\langle b_2, R_2, S_2 \rangle$
to $\langle b, R, S\rangle$.

\noindent The cases $f \in \Sigma_{\POP_{12}}$,
$f \in \Sigma_{\POP_{21}}$,
$f \in \Sigma_{\POP_{22}}$ are similar.


The above constructions
ensure the three invariants stated above, 
after the definition of $Q^\deter$ and corresponding to the three components of these states.
It follows that $L(\A) = L(\A^\deter)$.
\cqfd
\end{proof}

\subsection{Closure Properties}
The tree automata with one memory of~\cite{ComonCortier05tcs}
are closed under union but not closed under intersection
and complement (even their version without constraints).
The visibility condition makes possible these closures for VTAM.
%
\begin{thm} \label{th:vtam-closure}
The class of tree languages of VTAM is closed under Boolean operations.
One can construct VTAM for union, intersection and complement of given VTAM languages
whose sizes are respectively linear, quadratic and exponential in the size of the initial VTAM.
\end{thm}
%
\begin{proof} 
Let 
$\A_1 =  (\Gamma_1, Q_1, Q_{\final,1}, \Delta_1)$ and
$\A_2 =  (\Gamma_2, Q_2, Q_{\final,2}, \Delta_2)$
be two VTAM on $\Sigma$.
We assume wlog that $Q_1$ and $Q_2$ are disjoint.

For the union of the languages of  $\A_1$ and $\A_2$,
we construct a VTAM $\A_\cup$ whose 
memory signature, state set, final state set and rules set are the union 
of the  respective 
memory signatures, state sets, final state sets and rules sets
of the two given VTAM. 
We have $L(\A_\cup) = L(\A_1) \cup L(\A_2)$.
\[ \A_\cup =  (\Gamma_1 \cup \Gamma_2, Q_1 \cup Q_2, Q_{\final,1} \cup Q_{\final,2}, \Delta_1 \cup \Delta_2)\]

For the intersection of the languages of  $\A_1$ and $\A_2$,
we construct a VTAM $\A_\cap$ whose 
memory signature, state set and final state set are the Cartesian product
of the respective  memory signatures, state sets and final state sets
of the two given VTAM. 
\[ \A_\cap =  (\Gamma_1 \times \Gamma_2, Q_1 \times Q_2, Q_{\final,1} \times Q_{\final,2}, \Delta_\cap) \]
The rule set $\Delta_\cap$ of the intersection VTAM $\A_\cap$ 
is obtained by "product" of  rules of the two given VTAM with same function symbols.
The product of rules means Cartesian products of the respective states and
memory symbols pushed or popped.
More precisely, $\Delta_\cap$ is the smallest set of rules such that:
\begin{enumerate}[$\bullet$]
\item if $\Delta_1$ contains 
$f \bigl(q_{11}(y_1), q_{12}(y_2)\bigr) \to q_1\bigl(h_1(y_1, y_2)\bigr)$
and $\Delta_2$ contains 
$f \bigl(q_{21}(y_1),  q_{22}(y_2)\bigr) \to q_2\bigl(h_2(y_1, y_2)\bigr)$,
for some $f \in \Sigma_{\PUSH}$, then $\Delta_\cap$ contains
$f \bigl(\langle q_{11}, q_{21}\rangle(y_1), \langle q_{12}, q_{22}\rangle(y_2)\bigr) \to 
  \langle q_1, q_2\rangle\bigl(\langle h_1, h_2\rangle(y_1, y_2)\bigr)$.
\item \sloppypar

if $\Delta_1$ contains $f\bigl(q_{11}(h_1(y_{11}, y_{12})\bigr),
q_{12}(y_2)\bigr) \to q_1(y_{11})$ and $\Delta_2$ contains
$f\bigl(q_{21}(h_2(y_{11}, y_{12})\bigr), q_{22}(y_2)\bigr) \to
q_2(y_{11})$ for some $f \in \Sigma_{\POP_{11}}$, then $\Delta_\cap$
contains $f\bigl(\langle q_{11}, q_{2,1}\rangle(\langle h_1,
h_2\rangle(y_{11}, y_{12})\bigr), \langle q_{12},
q_{2,2}\rangle(y_2)\bigr) \to \langle q_1, q_2\rangle(y_{11})$

\item 
similarly for $\POP_{12}$, $\POP_{21}$ and $\POP_{22}$ 
\item 
if $\Delta_1$ contains $f\bigl(q_{11}(y_1), q_{21}(y_2)\bigr) \to q_1(y_1)$
and $\Delta_2$ contains $f\bigl(q_{21}(y_1), q_{22}(y_2)\bigr) \to q_2(y_1)$
for some $f \in \Sigma_{\INT_1}$,
then $\Delta_\cap$ contains
$f\bigl(\langle q_{11}, q_{2,1}\rangle(y_1), \langle q_{12}, q_{2,2}\rangle(y_2)\bigr)  
  \to \langle q_1, q_2\rangle(y_1)$
\item 
and similarly for $\INT_2$, $\INT_0$.
\end{enumerate}
We have then $L(\A_\cap) = L(\A_1) \cap L(\A_2)$.
Note that the above product construction for $\A_\cap$ 
is possible only because the visibility condition
ensures that two rules with the same function symbol in left-side
will have the same form. Hence we can synchronize memory operations on 
the same symbols.

For the complement, we use the construction of Theorem~\ref{th:vtam-determinization}
and a completion (this operation preserves determinism), 
and take the complement of the final state set of the VTAM obtained.
\cqfd
\end{proof}

\subsection{Decision Problems}
Every VTAM is a particular case of tree automaton with one memory of~\cite{ComonCortier05tcs}.  
Since the emptiness problem (whether the language accepted is empty or not)
is decidable for this latter class, it  is also decidable for VTAM.
However, whereas this problem is EXPTIME-complete for the automata of~\cite{ComonCortier05tcs},
it is only PTIME for VTAM.
%
%

\begin{thm}  \label{th:vtam-emptiness}
The emptiness problem is PTIME-complete for VTAM.
\end{thm}
%
\begin{proof}
Assume given a VTAM $\A = (\Gamma, Q, Q_\final, \Delta)$.
By definition, for each state $q \in Q$, 
the language $L(\A, q)$ is empty iff the memory language $M(\A, q)$ is empty. 
For each state $q$, we introduce a predicate symbol $P_q$ and we construct
Horn clauses in such a way that $P_q(m)$ belongs to the least Herbrand model 
of this set of clauses, 
iff the configuration with state $q$ and memory $m$ is reachable by the automaton
(i.e. $m \in M(\A, q)$).

For such a construction (already given in \cite{ComonCortier05tcs}), we simply
forget the function symbol, associating to a transition rule 
$f(q_1(m_1), q_2(m_2)) \rightarrow q(m)$ the Horn clause
$P_{q_1}(m_1), P_{q_2}(m_2) \Rightarrow P_{q}(m)$. 
Then, according to the restrictions in Definition~\ref{def:vtam}, we get only Horn clauses
of one of the following forms:
\[
\begin{array}{rcl}
                                                   & \Rightarrow & P_q(c)\\
P_{q_1}(y_1), P_{q_2}(y_2) & \Rightarrow & P_q\bigl(h(y_1, y_2)\bigr)\\
P_{q_1}\bigl(h(y_{11}, y_{12})\bigr), P_{q_2}(y_2) & \Rightarrow &  P_q(y_{11})\\ 
P_{q_1}\bigl(h(y_{11}, y_{12})\bigr), P_{q_2}(y_2) & \Rightarrow &  P_q(y_{12})\\ 
P_{q_1}(\bot), P_{q_2}(y_2) & \Rightarrow &  P_q(\bot)\\ 
P_{q_1}(y_1), P_{q_2}(y_2) & \Rightarrow &  P_q(y_1) 
\end{array}
\]
where all the variables are distinct. 
Such clauses belong to the class $\mathcal{H}_3$ of~\cite{Nielson02sas}, for which
it is proved in~\cite{Nielson02sas} that emptiness is decidable in cubic time.
It follows that emptiness of VTAM is decidable in cubic time.

Hardness for PTIME follows from the PTIME-hardness of emptiness
of finite tree automata~\cite{tata}.
\cqfd
\end{proof}

Another proof relying on similar techniques, but for 
 a more general result, will be stated in Lemma~\ref{lem:2ways}
and can be found in~Appendix~\ref{app:2ways}. 

The \emph{universality} is the problem of deciding whether a given automaton recognizes 
all ground terms. \emph{Inclusion} refers to the problem of deciding the inclusion between 
the respective languages of two given automata.
\begin{cor}  \label{cor:vtam-universality}
The universality and inclusion problem are EXPTIME-complete for VTAM.
\end{cor}
\begin{proof}
A VTAM $\A$ is universal iff the language of its complement automaton 
$\overline{\A}$ is empty,
and $L(\A_1) \subseteq L(\A_2)$ iff $L(\A_1) \cap L(\overline{\A_2}) = \emptyset$.
With the bounds given in Theorem~\ref{th:vtam-closure}
these problems can be decided in EXPTIME for VTAM
(these operations require a determinization of a given VTAM first).

The EXPTIME-hardness 
follows from the corresponding property of finite tree automata (see \cite{tata}
for instance).\cqfd
\end{proof}

The \emph{membership} problem is, given a term $t$ and an automaton $\A$, 
to know whether $t$ is accepted by $\A$.
\begin{cor} 
The membership problem is decidable in PTIME for VTAM.
\end{cor}
\begin{proof}
Given a term $t$
we can build a VTAM $\A_t$ which recognizes exactly the language $\{ t \}$.
The intersection of $\A_t$ with the given VTAM $\A$ recognizes
a non empty language iff $t$ belongs to the language of $\A$.
\cqfd
\end{proof}

\section{Visibly Tree Automata with Memory and Constraints} \label{sec:vtamc}
In the late eighties, some models of tree recognizers were obtained by adding
equality and disequality constraints in transitions of tree automata.
They have been proposed  in order to solve problems with
term rewrite systems or constraints systems with non-linear
patterns (terms with multiple occurrences of the same variable).
The tree automata of~\cite{BogaertTison92} for instance
can perform equality and disequality tests between subterms
located at brother positions of the input term.

In the case of tree automata with memory, constraints
are applied to the memory contents.
Indeed, each bottom-up computation step
starts with two states and two memories
(and ends with one state and one memory), and therefore, 
it is possible to compare the contents of these two memories,
with respect to some binary relation.

We state first the general definition of visibly tree automata with constraints on memories (Section~\ref{sec:vtamc-def}),
then give sufficient conditions on the binary relation for the emptiness decidability (Section~\ref{sec:vtamc-decision})
and show that, if in general regular binary relations do not satisfy these conditions
(and indeed, the corresponding class of constrained VTAM has an undecidable emptiness problem,
Section~\ref{sec:constraint-regular})
some relevant examples do satisfy them.
In particular, we study in Section~\ref{sec:constraint-structural} the case of
VTAM with structural equality constraints.
They enjoy not only decision properties but also good closure properties.
Some relevant examples of tree languages recognized by constrained VTAM 
of this class are presented at the end of the section.


\subsection{Definitions} \label{sec:vtamc-def}
Assume given a fixed equivalence relation $R$ on $\T(\Gamma)$.
We consider now two new categories for the symbols of $\Sigma$:
$\INT^{R}_{1}$ and  $\INT^{R}_{2}$, 
in addition to the eight previous categories of page~\pageref{def:VTAM-partition}.
%
The new categories correspond to 
the constrained versions of the transition rules $\INT_{1}$ and $\INT_{2}$
presented in Figure~\ref{fig:categories-constraints}.
The constraint $y_1 \mathop{R} y_2$ in the two first rules of 
Figure~\ref{fig:categories-constraints} is called \emph{positive} and
the constraint $y_1 \mathop{\neg R} y_2$ in the two last rules 
is called \emph{negative}.

\begin{figure}
\[
\begin{array}{lcllclcl}
\INT^{R}_{1} & & f_9\bigl(q_1(y_1), & q_2(y_2)\bigr)  & \lrstep{y_1 \mathop{R} y_2}{} & q(y_1) 
  & \quad & f_9 \in \Sigma_{\INT^{R}_{1}}\\
\INT^{R}_{2} & & f_{10}\bigl(q_1(y_1), & q_2(y_2)\bigr)  & \lrstep{y_1 \mathop{R} y_2}{} & q(y_2)
  & \quad & f_{10} \in \Sigma_{\INT^{R}_{2}}\\
\INT^{R}_{1} & & f_{11}\bigl(q_1(y_1), & q_2(y_2)\bigr)  & \lrstep{y_1 \mathop{\neg R} y_2}{} & q(y_1)
  & \quad & f_{11} \in \Sigma_{\INT^{R}_{1}}\\
\INT^{R}_{2} & & f_{12}\bigl(q_1(y_1), & q_2(y_2)\bigr)  & \lrstep{y_1 \mathop{\neg R} y_2}{} & q(y_2)
  & \quad & f_{12} \in \Sigma_{\INT^{R}_{2}}
\end{array}
\]
\caption{New transition categories for \VTAMC{R}{\neg R}.}
\label{fig:categories-constraints}
\end{figure}

\sloppypar We shall not extend the rules $\PUSH$ and $\POP$ with constraints for some reasons explained in section \ref{sec:push=}.
%
A ground term $t$ rewrites to $s$ by a constrained rule 
$f\bigl(q_1(y_1),  q_2(y_2)\bigr)  \lrstep{y_1 \mathop{c} y_2}{} r$
(where $c$ is either $R$ or $\neg R$)
if there exists a position $p$ of $t$ and a substitution $\sigma $ 
such that $t|_p =\ell\sigma$, $y_1\sigma \mathop{c} y_2 \sigma$
and $s= t[r\sigma]_p$. 

For example, if $R$ is term equality, the transition is performed only when the 
memory contents are identical.

\begin{defi} \label{def:vtamc}
A \emph{visibly tree automaton with memory and constraints} (\VTAMC{R}{\neg R}) 
on a signature $\Sigma$
is a tuple $(\Gamma, R, Q, Q_\final, \Delta)$ where
$\Gamma$, $Q$, $Q_\final$ are defined as for TAM,
$R$ is an equivalence relation on $\T(\Gamma)$
and $\Delta$ is a set of rewrite rules in one of the  above categories:
$\PUSH$, 
$\POP_{11}$, 
$\POP_{12}$, 
$\POP_{21}$, 
$\POP_{22}$, 
$\INT_0$,
$\INT_1$,
$\INT_2$,
$\INT^{R}_{1}$,
$\INT^{R}_{2}$.
\end{defi}
We let \VTAMC{R}{} be the subclass of \VTAMC{R}{\neg R}
with positive constraints only.
%
The acceptance of terms of $\T(\Sigma)$ and languages of term and memories
are defined and denoted as in Section~\ref{sec:vtam-definition}.

\bigskip
The definition of \emph{complete} \VTAMC{R}{\neg R}  is the same as for VTAM.
As for VTAM, every   \VTAMC{R}{\neg R} can be completed (with a polynomial overhead) 
by the addition of a trash state $q_\bot$. 
The only subtle difference concerns the constrained rules:
for every $f_9 \in \INT_1^R$ and every states $q_1, q_2$, 
\begin{enumerate}[$\bullet$]
\item if there is a rule $f_9\bigl(q_1(y_1),  q_2(y_2)\bigr) \lrstep{y_1 \mathop{R} y_2}{} q(y_1)$
and no rule of the form $f_9\bigl(q_1(y_1),  q_2(y_2)\bigr) \lrstep{y_1 \mathop{\neg R} y_2}{} q'(y_1)$,
then we add $f_9\bigl(q_1(y_1),  q_2(y_2)\bigr) \lrstep{y_1 \mathop{\neg R} y_2}{} q_\bot(y_1)$,
\item if there is a rule $f_9\bigl(q_1(y_1),  q_2(y_2)\bigr) \lrstep{y_1 \mathop{\neg R} y_2}{} q(y_1)$
and no rule of the form $f_9\bigl(q_1(y_1),  q_2(y_2)\bigr) \lrstep{y_1 \mathop{R} y_2}{} q'(y_1)$,
then we add $f_9\bigl(q_1(y_1),  q_2(y_2)\bigr) \lrstep{y_1 \mathop{R} y_2}{} q_\bot(y_1)$,
\item if there is no rule of the form $f_9\bigl(q_1(y_1),  q_2(y_2)\bigr) \lrstep{y_1 \mathop{R} y_2}{} q(y_1)$
or $f_9\bigl(q_1(y_1),  q_2(y_2)\bigr) \lrstep{y_1 \mathop{\neg R} y_2}{} q'(y_1)$,
then we add 
$f_9\bigl(q_1(y_1),  q_2(y_2)\bigr) \lrstep{y_1 \mathop{R} y_2}{} q_\bot(y_1)$ and
$f_9\bigl(q_1(y_1),  q_2(y_2)\bigr) \lrstep{y_1 \mathop{\neg R} y_2}{} q_\bot(y_1)$.
\end{enumerate}

The definition of \emph{deterministic} \VTAMC{R}{\neg R} 
is based on the same conditions as for VTAM for the function symbols in categories
of $\PUSH_0$, $\PUSH$,  $\POP_{11}$,  \ldots, $\POP_{22}$,  $\INT_1$, $\INT_2$.
For the function symbols of 
$\INT^{R}_{1}$, $\INT^{R}_{2}$, 
we have the following condition: 
for all $f \in \Sigma_{\INT^{R}_{1}} \cup \Sigma_{\INT^{R}_{2}}$ 
	for all $q_1, q_2 \in Q$, 
	there are at most two rules in $\Delta$ with left-member $f\bigl(q_1(y_1), q_2(y_2)\bigr)$, 
    and if there are two, one has a positive constraint and the other has a negative constraint.

We will see in  Section~\ref{sec:constraint-equality}
a subclass of \VTAMC{R}{\neg R} 
that can be determinized (when $R$ is structural equality)
and another one that cannot (when $R$ is syntactic equality).


\subsection{Sufficient Conditions for Emptiness Decision}  
\label{sec:vtamc-decision}
We propose here a generic theorem ensuring emptiness decision for \VTAMC{R}{\neg R}. 
The idea of this theorem is that under some condition on $R$, 
the transition rules with negative constraints can be eliminated.

\begin{thm} \label{th:vtamc-emptiness}
Let $R$ be an equivalence relation satisfying these two properties:
\begin{enumerate}[\em i.]
\item for every automaton $\A$ of \VTAMC{R}{} and for every state $q$ of $\A$,
	the memory language $M(\A, q)$ is effectively a regular tree language,
\item 
for every term $m \in \T(\Gamma)$, 
the cardinality of the equivalence class of $m$ for $R$ is finite and
and its elements can be enumerated.
\end{enumerate}
Then the emptiness problem is decidable for \VTAMC{R}{\neg R}.
\end{thm}
\begin{proof}
The proof relies on the following Lemma~\ref{lem:positif}
which states that the negative constraints in \VTAMC{R}{\neg R}
can be eliminated, while preserving the memory languages.
The elimination can be done thanks to the
 condition \textit{ii} , 
by  replacement of the rules of
$\INT^{\neg R}_{1}$ and $\INT^{\neg R}_{2}$
by rules of $\INT^{R}_{1}$ and $\INT^{R}_{2}$.

Next, we can use \textit{i} in order to decide emptiness for
the \VTAMC{R}{} obtained by elimination of negative constraints.
Indeed, for all states $q$ of $\A$, 
by definition,  $L(\A, q)$ is empty iff $M(\A, q)$ is empty.\cqfd
\end{proof}

\begin{lem} \label{lem:positif}
Let $R$ satisfy the hypotheses $i$ and $ii$ of Theorem~\ref{th:vtamc-emptiness}, 
and let $\A = (\Gamma, R, Q, Q_\final, \Delta)$ be a  \VTAMC{R}{\neg R}.
There exists a  \VTAMC{R}{} $A^+ =  (\Gamma, R, Q^+, Q_\final, \Delta^+)$ 
such that $Q \subseteq Q^+$,
and for each $q \in Q$, $M(\A^+, q) = M(\A, q)$.
\end{lem}
\begin{proof}
The construction of $A^+$ is by induction on the number $n$ of rules with negative constraints
in $\Delta$ 
and uses the bound on the size of equivalence classes, condition \emph{ii} of the theorem.

\noindent The result is immediate if $n = 0$.

\noindent We assume that the result is true for $n -1$ rules, and show that we can get
rid of a rule of $\Delta$ with negative constraints (and replace it with rules unconstrained or with positive constraints).
Let us consider one such rule:
\begin{equation}
f\bigl(q_1(y_1), q_2(y_2)\bigr) \lrstep{y_1 \mathop{\neg R} y_2}{} q(y_1)
\label{eq:neg-rule}
\end{equation}

We show that, under the induction hypothesis, we have the following lemma
which will be used below in order to get rid of the rule~(\ref{eq:neg-rule}).
\begin{lem} \label{lem:witness}
Given $m_1, \ldots, m_k \in M(\A, q_2)$, it is effectively decidable
whether $M(\A, q_2) \setminus \{ m_1, \ldots, m_k \}$ is empty or not
and, in case it is not empty, we can effectively build a $m_{k+1}$ in this
set.
\end{lem}
\begin{proof}
Let $[ m_i ]_R$ denote the equivalence class of $m_i$.
By condition \emph{ii}, every $[ m_i ]_R$ is finite, hence
for each $i \leq k$, 
we can build a VTAM $\A_i$ with a state $p_i$ such that
$M(\A_i, p_i)$ is the complement of $[ m_i ]_R$.
We add all the rules of $\A_i$ to $\A$, obtaining $\A'$
(we assume that the state sets of $\A_1,\ldots, \A_k, \A$ are disjoint, and
that the states of $\A_1,\ldots, \A_k$ are not final in $\A'$).

Since $R$ is an equivalence relation, we have:
\[ 
y_1 \mathop{\neg R} m_i
\mbox{~iff~}
y_1 \notin [ m_i ]_R
\mbox{~iff~}
\exists y_2 \notin [ m_i ]_R,\ y_1 \mathop{R} y_2
\]
Hence, if $y_2 = m_i$ is a witness for the rule (\ref{eq:neg-rule}),
then we can apply instead a rule:
\begin{equation}
f\bigl(q_1(y_1), p_i(y_2)\bigr) \lrstep{y_1 \mathop{R} y_2}{} q(y_1)
\label{eq:pos-rule}
\end{equation}
Then we add to $\A'$ the rules~(\ref{eq:pos-rule}) as above and obtain $\A''$.
It can be shown that $M(\A'', q_2) = M(\A, q_2)$.

Let $m_{k+1}$ be a term of $M(\A'', q_2) \setminus \{ m_1, \ldots, m_k \}$ of minimal size (if one exists).
This term $m_{k+1}$ can be created in a run of $\A''$ which does not use the rule~(\ref{eq:neg-rule}).
Otherwise, the witness for $y_2$ in the application of this rule
would be a term of $M(\A'', q_2) \setminus \{ m_1, \ldots, m_k \}$ smaller than $m_{k+1}$
(it cannot be one of $ \{ m_1, \ldots, m_k \}$ because for these particular values of $y_2$, 
we assume the application of~(\ref{eq:pos-rule})).
It follows that $m_{k+1} \in M(\A'' \setminus (\ref{eq:neg-rule}), q_2)$.
This automaton $\A_1 = \A'' \setminus (\ref{eq:neg-rule})$ has $n-1$ rules with negative constraints. 
Hence, by induction hypothesis, there is a  \VTAMC{R}{} 
$\A_1^+$ with $m_{k+1}$ in its memory language $M(\A_1^+, q_2)$. 
By condition \emph{i}, this language is regular
and we can build $m_{k+1}$ from a TA for this language.\cqfd
\end{proof}

Now, let us come back to the proof that  we can replace rule~(\ref{eq:neg-rule}), 
while preserving the memory languages.

\noindent If $M(\A, q_2) = \emptyset$ (which can be effectively
decided according to lemma \ref{lem:witness}) then the rule~(\ref{eq:neg-rule}) is useless
and can be removed from $\A$ without changing its memory language.
Note that the condition $M(\A, q_2) = \emptyset$ is decidable
because by hypothesis $i$, $M(\A, q_2)$ is regular.

\noindent Otherwise, let $m_1 \in M(\A, q_2)$ be built with Lemma~\ref{lem:witness}
and let $N_1$ be 
the cardinal of the equivalence class $[ m_1 ]_R$.
We apply $N_1$ times the construction of Lemma~\ref{lem:witness}. 
There are three cases:
\begin{enumerate}[(1)]

\item if we find more than $N_1$ terms in $M(\A, q_2)$, then one of them, 
say $m_k$ is not in $[ m_1 ]_R$.
Then~(\ref{eq:neg-rule}) is useless for the point of view of memory languages:
whatever value for $y_1$, we know a $y_2 \in M(\A, q_2)$ which permits to fire the rule.
Indeed, if $y_1 \in [ m_1 ]_R$, then we can choose $y_2 = m_k$, 
and otherwise we choose  $y_2 = m_1$.
Hence (\ref{eq:neg-rule}) can be replaced without changing the memory language 
by:
\begin{equation}
f\bigl(q_1(y_1), q_0(y_2)\bigr) \lrstep{}{} q(y_1)
\end{equation}
where $q_0$ is any state of $\A$ such that $M(\A, q_0) \neq \emptyset$.
We can then apply the induction hypothesis to the  \VTAMC{R}{\neg R} obtained.

\item if we find less than $N_1$ terms in $M(\A, q_2)$, but one is not in $[ m_1 ]_R$.
The case is the same as above.

\item if we find less than $N_1$ terms in $M(\A, q_2)$, all in $[ m_1 ]_R$, 
it means that one of the applications of Lemma~\ref{lem:witness} was not successful,
and hence that we have found all the terms of $M(\A, q_2)$.
It follows that the rule~(\ref{eq:neg-rule}) can be fired iff $y_1 \notin [m_1]_R$,
i.e. there exists $y_2 \notin [m_1]_R$ such that $y_1 R y_2$.
Hence, we can replace~(\ref{eq:neg-rule})  by
\[ f\bigl(q_1(y_1), p_1(y_2)\bigr) \lrstep{y_1 \mathop{R} y_2}{} q(y_1). \]
Then we can apply the induction hypothesis.\cqfd
\end{enumerate}
\end{proof}
We present in Section~\ref{sec:constraint-equality}
two examples of relations satisfying \textit{i.} and \textit{ii}.

\subsection{Regular Tree Relations} 
\label{sec:constraint-regular}
We first consider the general case of \VTAMC{R}{\neg R}
where the equivalence $R$ 
is based on an arbitrary regular  binary relation on $\T(\Gamma)$.
By regular binary relation, we mean a set of pairs of ground terms 
accepted by a tree automaton computing simultaneously in both 
terms of the pair. More formally, we use a coding of a pair
of terms of $\T(\Sigma)$ into a term of 
$\T\bigl((\Sigma \cup \{ \bot \})^2\bigr)$, where $\bot$ is a new
constant symbol (not in $\Sigma$).
This coding is defined recursively by:
\begin{enumerate}[$\bullet$] 
\item $\otimes: \T(\Sigma) \cup \{ \bot \} \times \T(\Sigma) \cup \{ \bot \} \to \T\bigl((\Sigma \cup \{ \bot \})^2\bigr)$
\item for all $a, b \in \Sigma_0 \cup \{ \bot \}$, $a \otimes b := \langle a, b \rangle$,
\item for all $a \in \Sigma_0 \cup \bot$, $f \in \Sigma_2$, 
$t_1, t_2 \in \T(\Sigma)$,
$f(t_1, t_2) \otimes a := \langle f, a \rangle(t_1 \otimes \bot, t_2 \otimes \bot)$
$a \otimes f(t_1, t_2) := \langle a, f \rangle(\bot \otimes t_1,  \bot \otimes t_2)$,
\item for all $f,g \in \Sigma_2$, $s_1, s_2, t_1, t_2 \in \T(\Sigma)$,
$f(s_1, s_2) \otimes g(t_1, t_2) := \langle f, g \rangle(s_1 \otimes t_1, s_2 \otimes t_2)$.
\end{enumerate} 
Then, a binary relation $R \subseteq \T(\Sigma) \times \T(\Sigma)$ is called regular
iff the set $\{ s \otimes t \bigm| (s, t) \in R \}$ is regular.
The above coding of pairs is unrelated to the product 
used in Theorem~\ref{th:vtam-closure}.

\begin{thm} \label{th:vtamr-membership}
The membership problem for \VTAMC{R}{\neg R} 
is NP-complete
when $R$ is a regular binary relation.
\end{thm}
\begin{proof}
Assume given a ground term $t \in \T(\Sigma)$ and a \VTAMC{R}{\neg R} 
$\A = (\Gamma, R, Q, Q_\final, \Delta)$.
Because of the visibly condition, for every subterm $s$ of $t$, 
we can compute in polynomial time in the size of $s$
the shape  denoted $\mathit{struct}(s)$, which is an abstraction
of the memory reached when $\A$ runs on $s$. 
More precisely, $\mathit{struct}(s)$ is an unlabeled tree, and every
possible content of memory $m$ reachable by $\A$ 
in a computation $s \lrstep{*}{\Delta} q(m)$ is obtained by a labeling
of the nodes of $\mathit{struct}(s)$ with symbols of $\Gamma$.
Note that for all subterm $s$, the size of $\mathit{struct}(s)$ is smaller than the size of $t$.

Let us guess a decoration of every node of $t$ with a state of $Q$ and
a labeling of $\mathit{struct}(s)$ (where $s$ is the subterm of $t$ at the given node),
such that the root of $t$ is decorated with a final state of $Q_\final$.
We can check in polynomial time whether this decoration
represents a run of $\A$ on $t$ or not.

The NP-hardness is a consequence of Theorem~\ref{th:vtam=-membership}, 
which applies to the particular case where $R$ is the syntactic equality between terms.\cqfd
\end{proof}
Note that the NP algorithm works with every equivalence $R$ based on a regular relation,
but the the NP-hardness concerns only some cases of such relations.
For instance, in Section~\ref{sec:constraint-equality},
we will see one example of relation for which membership 
is NP-hard and another example for which it is in PTIME.

\bigskip
The class of \VTAMC{R}{\neg R} when $R$ is a binary regular tree relation
constitutes a nice and uniform framework.
Note however the condition \emph{ii} of Theorem~\ref{th:vtamc-emptiness} 
is not always true in this case.
Actually, this class is too expressive.
%
\begin{thm}  \label{th:vtamc-regular-undec}
Given a regular binary relation $R$ and an automaton $\A$ in  \VTAMC{R}{},
the emptiness of $L(\A)$ is undecidable. 
\end{thm}
\begin{proof} 
We reduce the blank accepting problem for a deterministic Turing machine $\M$.
We encode configurations of $\M$ 
as "right-combs" (binary trees) built with the tape and state symbols of $\M$,
in $\Sigma_\PUSH$ (hence binary)  
and a constant symbol $\varepsilon$ in $\Sigma_{\INT_0}$.
Let $R$ be the regular relation which accepts all the pairs of configurations 
$c \otimes c'$ such that $c'$ is a successor of $c$ by $\M$.
A sequence of configurations $c_0 c_1\ldots c_n$ (with $n \geq 1$) 
is  encoded as a tree $t = f(c_0(f(c_1,\ldots f(c_{n-1}, c_n)))$,
where $f$ is a binary symbol of  $\Sigma_{\INT^R_1}$.

We construct a \VTAMC{R}{}  $\A$ which accepts exactly the term-representations $t$
of computation sequences of $\M$ starting with 
the initial configuration $c_0$ of $\M$ 
and ending with a final configuration $c_n$ with blank tape.
Following the type of the function symbols,  the rules of $\A$ will
\begin{enumerate}[$\bullet$]
\item push all the symbols read in subterms of $t$ 
corresponding to configurations,
\item  compare, with $R$,  $c_i$ and $c_{i+1}$ 
(the memory contents in respectively the left and right branches)
and store $c_i$ in the memory,
with a transition applied at the top of a subterm $f(c_i, f(c_{i+1}, \ldots))$.
\end{enumerate}

This way, $\A$ checks that successive configurations in $t$ 
correspond to transitions of $\M$, hence that the language of
$\A$ is not empty iff $\M$ accepts the initial configuration $c_0$.~\cqfd
\end{proof}


\subsection{Syntactic and Structural Equality and Disequality Constraints} \label{sec:constraint-equality}
We present now two examples of relations satisfying the conditions of  Theorem~\ref{th:vtamc-emptiness}:
syntactic and structural term equality.
The satisfaction of condition \emph{i} will be proved with the help of the following crux Lemma.

\begin{lem} \label{lem:2ways}
Let $R$ be a regular binary relation defined by a TA whose state set is 
$\bigl\{ R_i \bigm| i = \{ 1..n\} \bigr\}$ and such that 
\( \forall i, j\, \exists k,l,\: \forall x,y,z.\; x R_i y \wedge y R_j  z \Leftrightarrow x R_k y \wedge x R_l z \).

\noindent Let $\A = (\Gamma, R, Q, Q_\final, \Delta)$ 
be a tree automaton with memory and constraints (not necessarily visibly).
Then it is possible to compute in exponential time a finite
tree automaton $\A'$, such that, for every state $q\in Q$, the language
$M(\A,q)$ is the language accepted in some state of~$\A'$.
\end{lem}
\begin{proof} (Sketch)
To prove this lemma, we first observe that the $M(\A,q)$ (for $q\in Q$) are actually
the least sets that satisfies the following conditions (we assume here for simplicity that the non-constant symbols are binary and display only some of
the implications; the others can be easily guessed):
\begin{center}
\begin{tabular}{rl}
$\forall x,y,z.$ &
$x \in M(\A,q_1),y \in M(\A,q_2)) \Rightarrow g(x,y) \in M(\A,q)$\\
& \multicolumn{1}{r}{%
if there is a rule $f(q_1(x_1),q_2(x_2)) \rightarrow q(g(x_1,x_2))$}\\[1mm]
& $g(x,y)\in M(\A,q_1), z \in M(\A,q_2) \Rightarrow x \in M(\A,q)$\\ 
& \multicolumn{1}{r}{%
if there is a rule $f(q_1(g(x,y),q_2(z)) \rightarrow q(x)$}\\[1mm]
& $x\in M(\A,q_1), y\in M(\A, q_2), R(x,y) \Rightarrow x \in M(\A,q)$\\ 
& \multicolumn{1}{r}{%
if there is a rule $f(q_1(x),q_2(y)) \lrstep{x R y}{} q(x)$}\\
& $\cdots$
\end{tabular}
\end{center}


In terms of automata, this means that $M(\A,q)$ is a language recognized
by a two-way alternating tree automaton with regular binary constraints.
In other words, such languages are the least Herbrand model of a set of clauses
of the form

\[
\begin{array}{rcll}
Q_1(y_1), Q_2(y_2), R(y_1,y_2) & \Rightarrow & Q_3(y_1) & \mbox{  $\INT_1,\INT_2$}\\
 Q_1(y_1), Q_2(y_2) & \Rightarrow & Q_3(f(y_1,y_2)) & \mbox{  $\PUSH$}\\
 & \Rightarrow & Q_1(a) & \mbox{  $\INT_0$}\\
 Q_1(f(y_1,y_2)), Q_2(y_3) & \Rightarrow & Q_3(y_1) & \mbox{  $\POP_{11}, \POP_{21}$}\\
 Q_1(f(y_1,y_2)), Q_2(y_3) & \Rightarrow & Q_3(y_2) &\mbox{  $\POP_{12}, \POP_{22}$}\\
\end{array}
\]

The lemma then shows that languages that are recognized by two-way
alternating tree automata with some particular regular constraints, are
also recognized by a finite tree automaton. 
This 
corresponds to classical reductions of two-way automata
to one-way automata (see e.g \cite{tata}, chapter 7, \cite{goubault06tata}, or
\cite{FruhwirthShapiroVardiYardeni91,charatonik97lics} for the first relevant references).
The idea of the reduction is to find shortcuts: moving up and down yields
a move at the same level. Add such shortcuts as new rules, until getting a
``complete set''. Then only keep the non-redundant rules: this yields
a finite tree automaton. Such a procedure relies on the definitions of ordered
strategies, redundancy and 
saturation (aka complete sets), which are classical notions in automated
first-order theorem proving \cite{goubault06tata,bachmair01hb,nieuwenhuis01hb}. 
Indeed, formally, a ``shortcut'' must be a formula, which allows for
smaller proofs than the proof using the two original rules. A \emph{saturated
set} corresponds to a set of formulas whose all shortcuts are already in the
set. 

The advantage of the clausal formalism is to
enable an easy representation of the above shortcuts, as intermediary steps.
Such shortcuts are clauses, but are not automata rules. 
Second, we may rely on completeness results for Horn clauses.

That is why, only for the proof of this lemma, which follows and extend
the classical proofs adding some regular constraints, we switch to
a first-order logic formalization.
The complete proof can be found in Appendix~\ref{app:2ways}. 
As in the classical proofs, we saturate the set of clauses  by resolution with selection and eager splitting.
This saturation terminates, and the set of clauses corresponding to finite tree
automata transitions in the saturated set recognizes the language $M(\A, q)$,
which is therefore regular.
\cqfd
\end{proof}

The condition on $R$ in the lemma allows to break chains such as
 $\exists x_1,\ldots,x_n. x R x_1 \wedge x_1 R x_2 \wedge \cdots \wedge x_n R y \wedge P(x,y)$, which would be a source of non-termination in the saturation
procedure.  We may indeed replace such chains by 
$\exists x_1,\ldots,x_n. x R_1 x_1 \wedge x R_2 x_2 \wedge \ldots \wedge x R_n x_n\wedge x R_0 y \wedge P(x,y)$, which can again be simplified into 
$\exists x_1. x S x_1 \wedge x R_0 y \wedge P(x,y)$ where $S$ is the intersection of $R_1,\ldots,R_n$. Possible such intersections range in a finite set
as the relation $R$ is regular and the $R_i$s are states of the automaton accepting $R$. 

Finally note that finding $k,l$ in the lemma's assumption can always be
performed in an effective way since $R$ is regular.

\subsubsection{Syntactic Constraints.}
We first apply Lemma~\ref{lem:2ways} to the class \VTAMC{=}{\neq} 
where $=$ denotes the equality between ground terms made of memory symbols.
Note that it is a particular case of constrained \VTAMC{R}{\neg R} of the 
above section~\ref{sec:constraint-regular}, since the term equality is a regular relation.
The automata of the subclass with positive constraints only, \VTAMC{=}{},
are particular cases of tree automata with one memory  of~\cite{ComonCortier05tcs},
and have therefore a decidable emptiness problem.
We show below that \VTAMC{=}{\neq} fulfills the hypotheses of 
Theorem~\ref{th:vtamc-emptiness},
and hence that the emptiness is also decidable for the whole class.

We can first verify that the relation $=$ checks the hypothesis of 
Lemma~\ref{lem:2ways},
hence the condition \emph{i} of Theorem~\ref{th:vtamc-emptiness}.
%
Moreover, the relation $=$ obviously also checks the condition \emph{ii} of 
Theorem~\ref{th:vtamc-emptiness}.
\begin{cor} \label{th:vtameq-emptiness}
The emptiness problem is decidable for \VTAMC{=}{\neq}.
\end{cor}
A careful analysis of the proof of Theorem~\ref{th:vtamc-emptiness}
permits to conclude to an EXPTIME complexity for this problem with \VTAMC{=}{\neq}.

\begin{thm} \label{th:vtam=-membership}
The membership problem is NP-complete for \VTAMC{=}{\neq}.
\end{thm}
\begin{proof}
An NP algorithm is given in the proof of Theorem~\ref{th:vtamr-membership}.
For the NP-hardness, we use a logspace reduction of 3-SAT.
Let us consider an instance of 3-SAT with $n$ propositional variables
$X_1,\ldots, X_n$ and a conjunction of $m$ clauses:
\[ \bigwedge_{i = 1}^{m} ( \alpha_{i, 1} \vee \alpha_{i, 2} \vee \alpha_{i, 3} ) \] 
where every $\alpha_{i, j}$ is either a variable $X_k$ ($k \leq n$) or a negation
of variable $\neg X_k$.
We assume wlog that every variable occurs at most once in a clause.

We consider an encoding $t$ of the given instance as a term over the signature $\Sigma$
containing the symbols: $X_1,\ldots, X_n$ (constants),
$\mathit{id}$, $\mathit{false}$, $\neg$ (unary) and $\wedge$ and $\vee$ (binary).
The encoding is: 
\[ t := C_\wedge \bigl[ C_\vee [ \delta_{1,1}(X_1),\ldots, \delta_{1,n}(X_n) ], \ldots, 
                                 C_\vee [ \delta_{m,1}(X_1),\ldots, \delta_{m,n}(X_n) ] \bigr] \]
where $C_\wedge$ (resp. $C_\vee$) is a context built solely with $\wedge$ 
(resp. $\vee$)
and where every $\delta_{i,j}$ is either:
\begin{enumerate}[$\bullet$]
\item $\delta_{i,j} = \mathit{id}$ (interpreted as the identity) if one of $\alpha_{i, 1}, \alpha_{i, 2}, \alpha_{i, 3}$ is $X_j$, 
\item $\delta_{i,j} = \neg$ if one of $\alpha_{i, 1}, \alpha_{i, 2}, \alpha_{i, 3}$ is $\neg X_j$, 
\item $\delta_{i,j} = \mathit{false}$ (interpreted as the constant function returning $\mathit{false}$) 
if $X_j$ does not occur in $\alpha_{i, 1}, \alpha_{i, 2}, \alpha_{i, 3}$.
\end{enumerate}

Now, let us partition the signature $\Sigma$ with:
$X_1,\ldots, X_n, \vee \in \PUSH$, 
$\mathit{id}, \mathit{false}, \neg \in \INT_1$ and
$\wedge \in \INT^=_1$;
and let consider the memory signature $\Gamma = \{0, 1, \vee \}$.
We construct now a \VTAMC{=}{}
$\A = (\Gamma, =, \{ q_0, q_1 \} , \{ q_1 \}, \Delta)$
whose transition will, intuitively:
\begin{enumerate}[$\bullet$]
\item guess an assignment for each constant symbol $X_k$ of $t$, 
by mean of a non-deterministic choice of one state $q_0$ or $q_1$,
\item compute the value of $t$ with these assignments,
\item push each tuple of assignment for each clause,
         in the contexts $C_\vee$,
\item check the coherence of assignments by means of equality tests
     between the tuples pushed, in the context $C_\wedge$.
\end{enumerate}
More formally, we have the following transitions in $\Delta$:
\[
\begin{array}{rclc}
%
X_i & \to & q_0(0) \\  
X_i & \to & q_1(1) 
& 
i \leq n
\\[1mm]
\mathit{id}(q_\varepsilon(y_1)) & \to & q_\varepsilon(y_1)   \\
\mathit{false}(q_\varepsilon(y_1)) & \to & q_0(y_1) \\
\neg(q_\varepsilon(y_1)) & \to & q_{1-\varepsilon}(y_1) 
& 
\mbox{with~} \varepsilon \in \{ 0, 1 \} \\[1mm]
\vee(q_{\varepsilon_1}(y_1), q_{\varepsilon_2}(y_2)) 
 & \to & q_{\varepsilon_1 \vee \varepsilon_2}(\vee(y_1, y_2))\\
\wedge(q_{\varepsilon_1}(y_1), q_{\varepsilon_2}(y_2)) 
 & \lrstep{y_1 = y_2}{} & q_{\varepsilon_1 \wedge \varepsilon_2}(y_1)
& \mbox{with~} \varepsilon_1, \varepsilon_2 \in \{ 0, 1 \}
\end{array}
\]
We can verify that the above \VTAMC{=}{} $\A$ 
recognizes $t$ iff the instance of 3-SAT has a solution.\cqfd
\end{proof}

\VTAMC{=}{\neq} is closed under union
(using the same construction as before)
but not under complementation. This is a consequence of the following Theorem.

\begin{thm} \label{th:vtam=universality}
The universality problem is undecidable for \VTAMC{=}{\neq}.
\end{thm}
\begin{proof}
We reduce the blank accepting problem for a deterministic Turing machine $\M$.
Like in the proof of Theorem~\ref{th:vtamc-regular-undec},
we encode \emph{configurations} of $\M$ 
as right-combs on a signature $\Sigma$ containing
the tape and state symbols of $\M$,
considered as binary symbols of $\Sigma_\PUSH$
and a constant symbol $\varepsilon$ in $\Sigma_{\PUSH}$.
A sequence of configurations $c_0, c_1,\ldots, c_n$ (with $n \geq 1$) 
is  encoded as a tree $t = f(c_n(f(c_{n-1},\ldots f(c_{0}, \varepsilon))))$,
where $f$ is a binary symbol of  $\Sigma_{\INT^=_1}$.
Such a tree is called a \emph{computation} of $\M$
if $c_0$ is the initial configuration, $c_n$ is a final configuration and
for all $0 \leq i < n$, $c_{i+1}$ is the successor of $c_i$ with $\M$.
Moreover, we assume that all the $c_i$ have the same length
(for this purpose we complete the 
representations of configurations with blank symbols).

We want to construct a \VTAMC{=}{\neq} $\A$ which recognizes 
exactly the terms which are \emph{not} computations of $\M$.
Hence, $\A$ recognizes all the terms of $\T(\Sigma)$ iff $\M$ does not accept
the initial blank configuration.

For the construction of $\A$, let us first observe
that we can associate to $\M$ a VTAM $\A_\Box$
which, while reading a configuration $c_i$,
will push on the memory its successor $c_{i+1}$.
The existence of such an automaton is guaranteed
by the first fact that for each regular binary relation $R$,
as defined in Section~\ref{sec:constraint-regular},
there exists a VTAM which, for each $(s, t) \in R$, 
will push $t$ while reading $s$,
and by the second fact that the language of $c_i \otimes c_{i+1}$,
hence the relation of successor configuration,
are regular.
Moreover, since only push operations are performed, 
we can ensure that $\A_\Box$ satisfies the visibly condition.
Let us note $q_\Box$ the final state (which is assumed unique wlog)
of the VTAM $\A_\Box$.
We also use the following VTAMs:
\begin{enumerate}[\ ]
\item {\hskip-8 pt$\A_\forall$:} a VTAM with (unique) final state $q_\forall$ which, 
while reading a configuration $c_i$ will push on the memory any configuration 
with same length as $c_i$, 
\item {\hskip-8 pt$\A_=$:} a VTAM with final state $q_=$ which, 
while reading a configuration $c_i$ will push $c_i$ on the memory,
\item {\hskip-8 pt$\A_B$:} a VTAM with final state $q_B$ which, 
while reading a configuration $c_i$ will push on the memory 
a configuration  with same length as $c_i$ 
and containing only blank symbols.
\end{enumerate}

\noindent The \VTAMC{=}{\neq} $\A$ is the union of the following automata:
\begin{enumerate}[\ ]
\item {\hskip-8 pt$A_1$:} a \VTAMC{=}{\neq}  recognizing the terms of $\T(\Sigma)$
which are not representations of sequences of configurations 
(malformed terms). Its language is actually a regular tree language.
\item {\hskip-8 pt$A_2$:} a \VTAMC{=}{\neq}   recognizing the sequences of configurations 
$f(c_n(f(c_{n-1},\ldots f(c_{0}, \varepsilon))))$ such that $c_0$ is not initial
or $c_n$ is not final. Again, this is a regular tree language.
\item {\hskip-8 pt$A_3$:} a \VTAMC{=}{\neq}  recognizing the sequences of configurations 
with two configurations of different lengths.
It contains the transitions rules of $\A_B$ and
the additional transitions described in  Figure~\ref{fig:different-length},
which perform this test.
\item {\hskip-8 pt$A_4$:} a \VTAMC{=}{\neq} recognizing the sequences of configurations 
$f(c_n(f(c_{n-1},\ldots f(c_{0}, \varepsilon))))$ such that all the $c_i$ have the same length
but there exists $0 \leq i < n$
such that  $c_{i+1}$ is not the successor of $c_i$ by $\M$.
This last \VTAMC{=}{\neq} contains 
the transitions of $\A_\Box$, $\A_\forall$, $\A_=$, 
and the additional transitions described in  Figure~\ref{fig:not-successor}.
\end{enumerate}

\begin{figure}
\[
\begin{array}{cc}
\begin{array}[t]{rcl}
\varepsilon & \to & q_\epsilon(\varepsilon) \\ 
f(q_B(y_1), q_\varepsilon(y_2)) & \lrstep{y_1 \neq y_2}{} & q(y_1) \\
f(q_B(y_1), q(y_2)) & \lrstep{y_1 = y_2}{} & q(y_1) 
\end{array}
&
\begin{array}[t]{rcl}
f(q_B(y_1), q(y_2)) & \lrstep{y_1 \neq y_2}{} & q_\final(y_1)\\
f(q_B(y_1), q_\final(y_2)) & \lrstep{y_1 = y_2}{} & q_\final(y_1)\\
f(q_B(y_1), q_\final(y_2)) & \lrstep{y_1 \neq y_2}{} & q_\final(y_1)
\end{array}
\end{array}
\]
\caption{The \VTAMC{=}{\neq} $\A_3$ in the proof of Theorem~\ref{th:vtam=universality}.}
\label{fig:different-length}
\end{figure}

\begin{figure}
\[
\begin{array}{ccc}
\varepsilon \lrstep{}{} q_\varepsilon(\varepsilon) 
&
\begin{array}[t]{rcl}
f(q_\forall(y_1), q_\varepsilon(y_2)) & \lrstep{y_1 \neq y_2}{} & q_\forall(y_1) \\
f(q_\forall(y_1), q_\forall(y_2)) & \lrstep{y_1 = y_2}{} & q_\forall(y_1) \\
f(q_\Box(y_1), q_\forall(y_2))   & \lrstep{y_1 = y_2}{}   & q_\Box(y_1)\\
\end{array}
&
\begin{array}[t]{rcl}
f(q_=(y_1), q_\Box(y_2)) & \lrstep{y_1 \neq y_2}{}    & q_\final(y_1) \\
f(q_\forall(y_1), q_\final(y_2)) & \lrstep{y_1 = y_2}{} & q_\final(y_1) \\
\end{array}
\end{array}
\]
\caption{The \VTAMC{=}{\neq} $\A_4$ in the proof of Theorem~\ref{th:vtam=universality}.}
\label{fig:not-successor}
\end{figure}

With the transition rules in Figure~\ref{fig:not-successor},
the automaton $\A_4$ guesses a $i < n$ and,
while reading each of the configurations $c_j$ with $j \leq i$, 
it pushes the successor configuration of $c_j$, say $c'_j$
(second column of figure \ref{fig:not-successor}).
Then, while reading $c_{i+1}$ $\A_4$ pushes $c_{i+1}$,
and it checks that $c'_i$ and $c_{i+1}$ differ.
After that, when reading each of the remaining configurations, 
$\A_4$ pushes $c_{i+1}$ (third column of figure \ref{fig:not-successor}).

The \VTAMC{=}{\neq} $\A_1$ to $\A_4$ cover all the cases
of term $\T(\Sigma)$ not being an 
accepting computation of $\M$ starting with the initial blank configuration.
Hence the language of their union $\A$ is $\T(\Sigma)$ iff $\M$ does not 
accept the initial blank configuration.
\cqfd
\end{proof}

\begin{cor} \label{cor:boolean-vtam=}
\VTAMC{=}{\neq} is not effectively closed under complementation.
\end{cor}
\begin{proof}
 It is a consequence of 
 Corollary~\ref{th:vtameq-emptiness} (emptiness decision)
 and 
 Theorem~\ref{th:vtam=universality}.%
\cqfd
\end{proof}

\subsubsection{Structural Constraints.}

Lemma~\ref{lem:2ways} applies also to another class \VTAMS{}, 
where $\equiv$ denotes structural equality of terms,
\label{sec:constraint-structural}
defined recursively  as the smallest equivalence relation on ground terms such that:
\begin{enumerate}[$\bullet$]
\item $a \equiv b$ for all $a$, $b$ of arity $0$,
\item $f(s_1, s_2) \equiv g(t_1, t_2)$ 
	if $s_1 \equiv t_1$ and $s_2 \equiv t_2$,
 	for all $f$, $g$ of arity $2$.
\end{enumerate}
Note that it is a regular relation, and that it satisfies
the hypothesis of Lemma~\ref{lem:2ways} and the condition \emph{ii} of 
Theorem~\ref{th:vtamc-emptiness}.
\begin{cor} \label{th:vtams-emptiness}
The emptiness problem is decidable for \VTAMS.
\end{cor}
Following the procedure in the proof of Theorem~\ref{th:vtamc-emptiness},
we obtain a 2-EXPTIME complexity for this problem and this class.

\bigskip
The crucial property of the relations $\equiv$ and $\not\equiv$ is that,
unlike the above class  \VTAMC{=}{\neq} or the general \VTAMC{R}{\neg R},
they ignore the labels of the contents of the memory.
They just care of the structure of these memory terms.
A benefit of this property of  \VTAMS{} is that the decision of the membership
problem drops to PTIME  for this class.

\begin{thm} \label{th:vtams-membership}
The membership problem is decidable in PTIME for \VTAMS.
\end{thm}
\begin{proof}
Let $\A =  (\Gamma, \equiv, Q, Q_\final, \Delta)$ be a \VTAMS{} on $\Sigma$ 
and let $t$ be a term in $\T(\Sigma)$.
Let $\mathit{sub}(t)$ be the set of subterms of $t$ and let us construct a VTAM 
$\A' = (\Gamma, \mathit{sub}(t) \times Q, \{ t \} \times Q_\final, \Delta')$ on $\Sigma'$ where
the symbols of $\Sigma'$ and $\Sigma$ are the same, and we assume that the symbols
in category $\INT_1^\equiv$ (resp. $\INT_2^\equiv$) in the partition of $\Sigma$
are in $\INT_1$ (resp. $\INT_2$) in the partition of $\Sigma'$.
The transitions of $\Delta'$ are obtained by the following transformation of the transitions of $\Delta$.
We only describe the construction for the cases $\INT_1$ and $\INT_1^\equiv$ with positive constraints.
The other cases are similar.
\begin{enumerate}[$\bullet$]
\item for every $f_7(q_1(y_1), q_2(y_2)) \to q(y_1) \in \Delta$, 
we add to $\Delta'$ all the transitions:
$f_7\bigl( \langle q_1,t_1\rangle(y_1), \langle q_2, t_2\rangle(y_2)\bigl) \to 
 \bigl\langle q, f(t_1,t_2)\bigr\rangle(y_1)$
such that $f(t_1,t_2) \in \mathit{sub}(t)$,
\item for every $f_9(q_1(y_1), q_2(y_2)) \lrstep{y_1 \equiv y_2}{} q(y_1) \in \Delta$, 
we add to $\Delta'$ all the transitions as above (in this case, $f_9$ is assumed a symbol of 
category $\INT_1$ in $\Sigma'$) 
such that moreover $\mathit{struct}(t_1) = \mathit{struct}(t_2)$, 
where $\mathit{struct}(s)$ is defined, like in the proof of Theorem~\ref{th:vtamr-membership},
as the shape (unlabeled tree) that will have the memory of $\A$ after $\A$ processed $s$.
\end{enumerate}
The VTAM $\A'$ can be computed in time $O(\|t\|^2 \times \|A \|)$.
It recognizes at most one term, $t$,  
and it recognizes $t$ iff $\A$ recognizes $t$.
Therefore, $t$ is recognized by $\A$ iff
the language of $\A'$ is not empty. 
This can be decided in PTIME
according to Theorem~\ref{th:vtam-emptiness}.\cqfd
\end{proof}

Even more interesting, 
the construction for determinization  of Section~\ref{sec:vtam-determinism} still works for \VTAMS.
\begin{thm} \label{th:vtams-determinisation}
For every \VTAMS{} $\A = (\Gamma, \equiv, Q, Q_\final, \Delta)$ there exists a deterministic 
\VTAMS{} $\A^\deter = (\Gamma^\deter, \equiv, Q^\deter, Q_\final^\deter, \Delta^\deter)$ 
such that $L(\A) = L(\A^\deter)$, 
where $| Q^\deter |$ and $| \Gamma^\deter |$ both are $O\bigl(2^{|Q|^2}\bigr)$.
\end{thm}
\begin{proof}
We use the same construction as in the proof of Theorem~\ref{th:vtam-determinization},
with a direct extension of the construction for $\INT$ to $\INT^\equiv$. 
The key property for handling constraints 
is that the structure of memory (hence the result
of the structural tests) is independent from the non-deterministic choices of the automaton.
With the visibility condition it only depends on the term read.
%
%
\cqfd
\end{proof}

\begin{thm} \label{th:vtams-closure}
The class of tree languages of \VTAMS{} is closed under Boolean operations.
One can construct \VTAMS{} for union, intersection and complement of given \VTAMS{} languages
whose sizes are respectively linear, quadratic and exponential in the size of the initial \VTAMS.
\end{thm}
\begin{proof}
We use the same constructions as in Theorem~\ref{th:vtam-closure} (VTAM) for union and intersection.
For the intersection, in the case of constrained rules
we can safely keep the constraints in product rules, 
thanks to the visibility condition 
(as the structure of memory only depends on the term read, 
see the proof of Theorem~\ref{th:vtams-determinisation}).
For instance, the product of the $\INT^\equiv_1$ rules
$f_9 \bigl(q_{11}(y_1), q_{12}(y_2)\bigr) \lrstep{y_1 \equiv y_2}{} q_1(y_1)$
and
$f_9 \bigl(q_{21}(y_1),  q_{22}(y_2)\bigr) \lrstep{y_1 \equiv y_2}{} q_1(y_1)$
is
$f_9 \bigl(\langle q_{11}, q_{21}\rangle(y_1), \langle q_{12}, q_{22}\rangle(y_2)\bigr) 
\lrstep{y_1 \equiv y_2}{}  \langle q_1, q_2\rangle(y_1)$.
The product of two $\INT^{\not\equiv}_1$ is constructed similarly.
We do not need to consider the product 
of a rule $\INT^\equiv_1$ with a rule $\INT^{\not\equiv}_1$,
and vice-versa,
because in this case the product is empty 
(no rule is added to the \VTAMS{} for intersection).
For the complementation, we use Theorem~\ref{th:vtams-determinisation}
and completion.
\cqfd
\end{proof}

\begin{cor}  \label{th:vtams-universality}
The universality and inclusion problems are decidable for \VTAMS.
\end{cor}
\begin{proof}

This is a consequence of 
Corollary~\ref{th:vtams-emptiness} and Theorem~\ref{th:vtams-closure}.
\cqfd
\end{proof}

\subsection{Constrained $\PUSH$ Transitions} 
\label{sec:push=}
Above, we always considered constraints in transitions with 
$\INT$ symbols only.
We did not consider a constrained extension of the rules $\PUSH$.
The main reason is that symbols of a new category $\PUSH^\equiv$,
which test two memories for structural equality and then push a symbol
on the top of them, permit us to construct a constrained VTAM $\A$
whose memory language $M(\A, q)$ is the set of well-balanced binary trees.
This language is not regular, whereas the base of 
our emptiness decision procedure is the result
(Theorem~\ref{th:vtamc-emptiness}, Lemma~\ref{lem:2ways})
of regularity of these languages for the classes considered.


\subsection{Contexts as Symbols and Signature Translations} 
\label{sec:contexts}
Before looking for some examples of \VTAMS{} languages, we show a "trick" 
that (seemingly) adds expressiveness to \VTAMS.
One symbol can perform either a $\PUSH$ or a $\POP$ operation, or make an $\INT$ transition (constrained or not), 
but it cannot combine several of these operations. 
Here, we propose a way to combine several operations in one symbol, 
and thus increase the expressiveness of \VTAMS, without losing the good properties of this class.

The trick is to replace symbols by \emph{contexts}. 
For instance a context $g_2(g_1(\cdot, \cdot), g_0)$ 
can replace a symbol of arity 2. 
Assume that $g_2$ is a $\PUSH$ symbol, $g_1$ is an $\INT_1$ symbol with test, 
and $g_0$ is an $\INT_0$ symbol. 
This context first performs a test on the memories of the sons, 
and then a $\PUSH$ operation on the memory kept by $g_1$ (and on the $\bot$ leaf created by $g_0$). 
Such a combination is normally not possible, and replacing symbols by contexts brings a lot of additional 
expressiveness.

Here is how we precisely proceed: we want to recognize a language (on a signature $\Sigma$)  with a VTAM, 
and we have then to choose the categories for each symbol of the signature ($\PUSH$, $\POP_{ij}$, $\INT_1^\equiv$, ...). 
As we will see in the examples below, it might be useful in practice to have some extra categories 
combining the powers of two or more  categories of \VTAMS.
We can do that still with \VTAMS, by mean of an encoding of the terms of $\T(\Sigma)$.
More precisely, we replace some symbols of the initial signature $\Sigma$ 
by contexts built with new 
symbols.
For instance, we replace a $g \in \Sigma$, which will perform the complex operation described above,
by the context $g_2(g_1(\cdot , \cdot), g_0)$. 
Then, we will have to ensure that the new symbols 
(in our example $g_0$, $g_1$ and $g_2$) are only used to form the 
contexts encoding the symbols of $\Sigma$.
This can easily be done with local information maintained in the state of the 
automaton. The set of well formed terms, built with new symbols organized
in allowed contexts, is a regular tree language.
We will call the \VTAMS{} signature obtained a \emph{translation} of the initial signature. If $L$ is a tree language on $\Sigma$, then $c(L)$ is the translation of $L$. 

In summary, we have shown here a general method for adding new  categories of symbols corresponding
to (relevant) combinations of operations of \VTAMS, and hence defining extensions of \VTAMS{} with
the same good properties as \VTAMS.
By \emph{relevant}, we mean that  some combinations are excluded,
like for instance, $\PUSH$ + constraint $\equiv$ at the same time (see paragraph above).
Such forbidden combination cannot be handled by our method.
With similar encodings, we can deal with symbols of arity bigger than 2, \emph{e.g.} 
$g(\cdot, \cdot, \cdot)$ can be replaced by $g_2(\cdot, g_1(\cdot ,\cdot))$.

Note however first that this encoding concerns the recognized tree, 
\emph{not the
memories}. For instance, it is not possible to systematically encode
the syntactic equality as structural equality (on memories) in this way.
And indeed, the decision results are drastically different in the two cases.

Also note that, even if $c(L)$ is accepted by a VTAM, which implies that
$\neg c(L)$ is also accepted by a VTAM, it may well be the case that
$c(\neg L)$ is not recognized by a VTAM. So, the above trick does not
show that we can extend our results to a wider class of tree languages.

\subsection{Some \VTAMS{} Languages} 
\label{sec:examples}
The regular tree languages and VPL are particular cases of  VTAM languages.
%
We present in this section some other examples of relevant tree languages
translatable, using the method of Section~\ref{sec:contexts}, into \VTAMS{} languages.

\subsubsection*{Well balanced binary trees}
The \VTAMS{} with memory signature $\{ f, \bot \}$,
state set $\{q, q_\final \}$, 
unique final state $q_\final$,
and whose rules follow accepts the (non-regular) 
language of well balanced binary trees build
with $g$ and $a$.

Here $a$ is a constant  in $\Sigma_{\INT_0}$,
and $g$ is in a new category, and is translated into
the context $g_2(g_1(\cdot , \cdot), g_0)$, where
$g_2 \in \Sigma_\PUSH$, $g_1 \in \Sigma_{\INT^\equiv_1}$, 
and $g_0 \in \Sigma_{\INT_0}$.

\[
\begin{array}{ccc}
\begin{array}[t]{rcl}
a & \to & q_\final(\bot) \\
g_0 & \to & q_0(\bot)
\end{array}
& & 
\begin{array}[t]{rcl}
g_1\bigl(q_\final(y_1), q_\final(y_2)\bigr) & \lrstep{y_1 \equiv y_2}{} & q(y_1)\\
g_2\bigl(q(y_1), q_0(y_2)\bigr) & \longrightarrow & q_\final\bigl(f(y_1, y_2)\bigr)
\end{array}
\end{array}
\]

\subsubsection*{Powerlists}
A powerlist~\cite{Misra94powerlist} is roughly a list of length $2^n$ (for $n \geq 0$)
whose elements are stored in the leaves of a balanced binary tree.
For instance, the elements may be integers represented in unary notation
with the unary successor symbol $s$ and the constant $0$, and 
the balanced binary tree on the top of them can be built with a binary symbol $g$.
This data structure has been used in~\cite{Misra94powerlist} to specify 
data-parallel algorithms based on divide-and-conquer strategy and recursion
(\emph{e.g.} Batcher's merge sort and fast Fourier transform).

It is easy following the above construction to characterize 
translations of powerlists with a \VTAMS.
We do not push on the "leaves", \emph{i.e.} on the elements
of the powerlist, and compute in the higher part (the complete binary tree)
as above.

Some equational properties of algebraic specifications
of powerlists have been studied in the context of automatic induction
theorem proving and sufficient completeness~\cite{Kapur97}.
Tree automata with constraints have been acknowledged as a very powerful
formalism in this context (see \emph{e.g.}~\cite{ComonJacquemard03}).
We therefore believe that a characterization of powerlists (and their complement language)
with \VTAMS{} is useful for the automated verification of algorithms on this data structure.

\subsubsection*{Red-black trees}
A red-black tree is a binary search tree following these properties:
\begin{enumerate}[(1)]
\item  
every node is either red or black,
\item  
the root node is black,
\item  
all the leaves are black,
\item 
if a node is red, then both its sons are black,
\item  
every path from the root to a leaf contains the same number of black nodes.
\end{enumerate}

The four first properties are local and can be checked with standard TA rules.
The fifth property make the language red-black trees not regular and we
need \VTAMS{} rules to recognize it.
It can be checked by pushing all the black nodes read.
We use for this purpose a symbol $\mathit{black} \in \Sigma_{\PUSH}$.

When a red node is read, the number of black nodes in both its
sons are checked to be equal (by a test $\equiv$ on the corresponding memories) 
and only one corresponding memory is kept.
This is done with a symbol $\mathit{red} \in \Sigma_{\INT_1^{\equiv}}$.

When a black node is read, the equality of number of black nodes in its
sons must also be tested, and a $\mathit{black}$ must moreover be pushed on the 
top of the memory kept.
It means that two operations must be combined. 
We can do that by defining an appropriate context with the method of Section~\ref{sec:contexts}.

In~\cite{Habermehl06tacas} a special class of tree automata is introduced
and used in a procedure for the verification of C programs which handle balanced tree data structures, 
like red-black tree. Based on the above example, we think that, following the same approach,
\VTAMS{} can also be used for similar purposes.


\section{Visibly Tree Automata with Memory and Structural Constraints and Bogaert-Tison Constraints} 
\label{sec:bogaert-equality}   \label{sec:BT}

In Section~\ref{sec:vtamc}, we have only considered VTAM with constraints 
testing the \emph{memories} contents.
In this section,  we go a bit further and add to \VTAMC{R}{\neg R} some
Bogaert-Tison constraints~\cite{BogaertTison92}, 
i.e. equality and disequality tests between \emph{brother} subterms in the term read by the automaton.

We consider two new categories for the symbols which we call $\BTINT_1$
and $\BTINT_2$, for "Bogaert-Tison Internal".
A transition with a symbol in one of these categories will make no test 
on the memory contents, but rather an equality or disequality
test between the brother subterms directly under the current position of computation.
In Figure~\ref{fig:categories-bogaert}, we describe the new transitions categories. 
We use the same notation as in~\cite{BogaertTison92} for the constraints.
Note that again, we only allow Bogaert-Tison constraints in internal rules.

\begin{figure}
\[
\begin{array}{lcllclcl}
\BTINT_1 & & f_{13}\bigl(q_1(y_1), & q_2(y_2)\bigr)  & \lrstep{1 = 2}{} & q(y_1)
 & \quad &  f_{13} \in \Sigma_{\BTINT_1}\\
\BTINT_2 & & f_{14}\bigl(q_1(y_1), & q_2(y_2)\bigr)  & \lrstep{1 = 2}{} & q(y_2)
 & \quad &  f_{14} \in \Sigma_{\BTINT_2}\\
\BTINT_1 & & f_{15}\bigl(q_1(y_1), & q_2(y_2)\bigr)  & \lrstep{1 \neq 2}{} & q(y_1)
 & \quad &  f_{15} \in \Sigma_{\BTINT_1}\\
\BTINT_2 & & f_{16}\bigl(q_1(y_1), & q_2(y_2)\bigr)  & \lrstep{1 \neq 2}{} & q(y_2)
 & \quad &  f_{16} \in \Sigma_{\BTINT_2}
\end{array}
\]
\caption{New transition categories for \BTVTAMC{R}{\neg R}.}
\label{fig:categories-bogaert}
\end{figure}

For instance, if  $f_{13}(t_1, t_2)$ is a subterm of the input tree, 
and if $t_1$ leads to $q_1(m_1)$, and $t_2$ to $q_2(m_2)$, 
then the transition rule  $f_{13}\bigl(q_1(y_1), q_2(y_2)\bigr) \lrstep{1 = 2}{} q(y_1)$, 
of type $\BTINT_{1}$ can be applied at this position \emph{iff} $t_1 = t_2$.

\begin{defi} \label{def:btvtam}
A \emph{visibly tree automaton with memory and constraints and Bogaert-Tison tests} (\BTVTAMC{R}{\neg R}) 
on a signature $\Sigma$
is a tuple $(\Gamma, R, Q, Q_\final, \Delta)$ where
$\Gamma$, $Q$, $Q_\final$ are defined as for TAM,
$R$ is an equivalence relation on $\T(\Gamma)$
and $\Delta$ is a set of rewrite rules in one of the  above categories:
$\PUSH$, 
$\POP_{11}$, 
$\POP_{12}$, 
$\POP_{21}$, 
$\POP_{22}$, 
$\INT_0$,
$\INT_1$,
$\INT_2$,
$\INT^{R}_1$,
$\INT^{R}_2$,
$\BTINT_1$,
$\BTINT_2$.
\end{defi}
The acceptance of terms of $\T(\Sigma)$ and languages of term and memories
are defined and denoted as in Section~\ref{sec:vtam-definition}.

\bigskip
The definition of \emph{complete} \BTVTAMC{R}{\neg R} is the same as before.
Every   \BTVTAMC{R}{\neg R} can be completed (with a polynomial overhead) 
by the addition of a trash state $q_\bot$
(the construction is similar to the one for \VTAMC{R}{\neg R} in Section~\ref{sec:vtamc-def}).

The definition of \emph{deterministic} \BTVTAMC{R}{\neg R} 
is based on the same conditions as for \VTAMC{R}{\neg R} 
for the function symbols in categories $\PUSH_0$, $\PUSH$,  $\POP_{11}$,  \ldots, $\POP_{22}$,  
$\INT_1$, $\INT_2$, $\INT^{R}_{1}$, $\INT^{R}_{2}$, 
and for the function symbols of $\BTINT_{1}$, $\BTINT_{2}$, 
we use the same kind of conditions as for $\INT^R_1$, $\INT^R_2$:
for all $f \in \Sigma_{\BTINT_{1}} \cup \Sigma_{\BTINT_{2}}$ 
	for all $q_1, q_2 \in Q$, 
	there are at most two rules in $\Delta$ with left-member $f\bigl(q_1(y_1), q_2(y_2)\bigr)$, 
    and if there are two, then their constraints have different signs.

\begin{thm} \label{th:bogaert-determinisation}
For every \BTVTAMC{\equiv}{\not\equiv} 
$\A = (\Gamma, \equiv, Q, Q_\final, \Delta)$ there exists a deterministic 
\BTVTAMC{\equiv}{\not\equiv} $\A^\deter = (\Gamma^\deter, \equiv, Q^\deter, Q_\final^\deter, \Delta^\deter)$ 
such that $L(\A) = L(\A^\deter)$, 
where $| Q^\deter |$ and $| \Gamma^\deter |$ both are $O\bigl(2^{|Q|^2}\bigr)$.
\end{thm}
\begin{proof}
We use, again, the same construction as in the proof of Theorem~\ref{th:vtam-determinization},
with a direct extension of the construction for $\INT$ to $\INT^\equiv$ 
and $\BTINT$. 
As mentioned in Theorem~\ref{th:vtams-determinisation},
the extension works for $\INT^{\equiv}$  because the results of the tests are independent 
from the non-deterministic choices of the automaton. 
For $\BTINT$ 
it is exactly the same (the brother terms are not changed by the automaton!).
\cqfd
\end{proof}

\begin{thm} \label{th:bogaert-closure}
The class of tree languages of \BTVTAMC{\equiv}{\not\equiv} is closed under Boolean operations.
\end{thm}

\begin{proof}
We use the same constructions as in Theorem~\ref{th:vtam-closure} for union and intersection.
For the intersection, as in Theorem~\ref{th:vtams-closure}, the constraints (even Bogaert-Tison tests) 
can be safely kept in product rules, thanks to the visibility condition.
For the complementation, we use Theorem~\ref{th:bogaert-determinisation} and complementation. 
\cqfd
\end{proof}

The proof of the following theorem 
follows the same idea as the proof for Bogaert-Tison automata~\cite{BogaertTison92},
but we need here to take care of the structural constraints on the memory contents.
A consequence is that the complexity of emptiness decision is much higher.
\begin{thm} \label{th:bogaert-emptiness}
The emptiness problem is decidable for \BTVTAMC{\equiv}{\not\equiv}.
\end{thm}
\begin{proof}
Let $\A$ be a \BTVTAMC{\equiv}{\not\equiv}. 
First we determinize it into $\A^\deter$ and assume  that $\A^\deter$ is also complete.
Then, we delete the rules \( \BTINT_{1} \) of the form:
\(f \bigl(q_1(y_1), q_2(y_2)\bigr) \lrstep{1 = 2}{} q(y_1)\).
with $q_1$ distinct from $q_2$ (idem for \( \BTINT_2 \) rules) 
because they can't be used (the automaton is deterministic so one term cannot lead to two different states). 

For the same reason, we change each rule \( \BTINT_1 \) 
of the form:
\(f \bigl(q_1(y_1), q_2(y_2)\bigr) \lrstep{1 \neq 2}{} q(y_1)\)
with $q_1$ distinct from $q_2$ (idem for \( \BTINT^{\neq}_{2} \) rules) 
into the same rule but without the disequality test:
\(f \bigl(q_1(y_1), q_2(y_2)\bigr) \to q(y_1) \).

We call the newly obtained automaton $\A^\new$. 
It is still deterministic and recognizes the same language as $\A^\deter$.
Actually, the careful reader may notice that $\A^\new$ is not a true \BTVTAMC{\equiv}{\not\equiv},
because some unconstrained rules may involve symbols in $\BTINT$ in this automaton.
However, it is just an intermediate step in the construction of another automaton $\A'$ below.

Now, we consider the remaining $\BTINT_1$ or $\BTINT_2$ rules
with negative Bogaert-Tison constraints, which are of the form: 
\(f \bigl(q_1(y_1), q_1(y_2)\bigr) \lrstep{1 \neq 2}{} q(y_1)\) (or  $q(y_2)$).
We denote them by $R_1, ..., R_i,..., R_N$, 
and denote by $q_i$ the state in the left member of $R_i$,  for each $i \leq N$.
We also denote the corresponding $\BTINT_1$ or $\BTINT_2$
rules by $S_1$,...., $S_i$,..., $S_N$. 
Note that, since $\A^\deter$ is deterministic and complete, 
we can associate to each rule of $\BTINT_i$, whose constraint is negative,
a unique rule of $\BTINT_i$  with a positive constraint and the same states in its left member. 
So, the state in the left member of $S_i$ is the same $q_i$ as for $R_i$.

It is important to notice that if a rule $R_i$ can effectively be used, 
then there must exist two distinct terms leading to the state $q_i$ (we will call them witnesses). 
If not, the rule can be removed.

So, our purpose is now to find, for each rule $R_i$, whether two witnesses exist or not. 
We let $\R$ be initially $\{ R_1, \ldots, R_N \}$.
Suppose that at least one $R_i$ rule can be used, and consider a run on a term $t$ that uses such a rule. 
We consider an innermost application of a rule $R_i$ in this run on a subterm $f(t_1, t_2)$. 
The run on $t_1$ and the run on $t_2$ both lead to the state $q_i$, without any use of an $R_j$ rule.

Let us remove all the $R_i$ rules from $\A^\new$, 
and we remove all the equality tests in the $S_i$ rules. 
Let $\A'$ be the resulting automaton. 
It is a deterministic \VTAMS{} 
(considering the symbols in $\BTINT$ as $\INT$ symbols in this new automaton), 
and each term in $L(\A',q_i)$  can be transformed 
(we will call it \emph{BT-transformation}) into a term in $L(\A^\new,q_i)$: 
each time we use a modified $S_i$ rule, for instance of type $\BTINT_1$, 
on a subtree $f(t_1,t_2)$, we replace $t_2$ with $t_1$ so that the equality test is satisfied 
(and the resulting memory is unchanged). 
Important: all the replacements must be performed bottom-up.

The proof of the emptiness decidability of 
\VTAMS{} (Corollary~\ref{th:vtams-emptiness}) is constructive, 
hence if we choose a reachable state $q_j$, 
we can find a term in  $L(\A',q_j)$ to this state, and then convert it into a witness. 
So, we can find a first witness $t_A\in L(\A^\new,q_j)$.

If no witness can be found, then all the $R_i$ rules are useless and we can definitely remove them all. 
Otherwise, we still need to find another witness, 
and if there is at least one such other witness, then one of them can be recognized without using a $R_i$ rule. 
We can construct a \VTAMS{} recognizing all the terms whose BT-transformation leads to $t_A$. 
To design it, we read $t_A$ top-down (knowing the state of $\A'$ at each node), 
and each time we see a subterm $f(t_1, t_2)$ to which a modified $S_i$ rule has to be applied, 
for instance a modified $\BTINT_1$ (resp. $\BTINT_2$) rule, 
the right (resp. left) son of $f$ only needs to be a term in $L(\A',q_i)$, 
and the left (resp. right) son of $f$ only needs to be BT-transformed  into $t_1$ (resp. $t_2$). 
Once this \VTAMS{} is constructed, we can combine it with $\A'$ in order to obtain a 
\VTAMS{} recognizing all the terms leading $\A'$ to $q_j$ (the state reached by $\A'$ on $t_A$) 
except the terms whose BT-transformation is $t_A$. 
Then we find another term in $L(\A',q_j)$ (if it exists) 
and its BT-transformation is not $t_A$: it is actually another witness $t_B$.

When we have two witnesses for a rule $R_j$, we remove it from $\R$, 
and we add this rule $R_j$ to $\A'$, but without the disequality test. 
The automaton $\A'$ keeps its good property: a term $t$ leading $\A'$ to some state $q$ can be BT-transformed 
into a term leading $\A^\new$ to state $q$: when we "meet" the use of a rule formerly in the set $\R$
on $f(t_1, t_1)$ during the bottom-up exploration of $t$, 
we replace the right (for a rule that was of type $\BTINT_1$ and with negative constraints) 
or the left son (otherwise) by a witness different from $t_1$,  so that the disequality test is satisfied.
Note that even if $t_1$ is a witness, we can do so because we have found two witnesses.

With the new rule in $\A'$ we look for 2 witnesses for some remaining $R_i$ rule.
Again, we can show that if a couple of witnesses exists, then at least one couple 
can be found without any use of the remaining $R_i$ rules.
When we find a first witness $t_A$ for a remaining rule $R_j$, 
we can find another one (if it exists) using approximately the same technique as previously: 
we read $t_A$ top-down, and when we see a rule formerly in $\R$, used on $f(t_1,t_2)$ 
(e.g. a rule formerly of type $\BTINT_1$ with a negative constraint), 
we just go on recursively, saying that the left son must be a term whose 
BT-transformation is $t_1$, and the right son must be either:
\begin{enumerate}[$\bullet$]
\item a term whose BT-transformation is $t_2$,
\item or, if our BT-transformation would change $f(t_1, t_1)$ into $f(t_1, t_2)$, 
a term whose BT-transformation is $t_1$.
\end{enumerate}
As previously, we construct a \VTAMS{}, fully using the Boolean closure of this class, 
that recognizes the terms in $L(\A',q_j)$ (the state reached by $\A'$ on $t_A$), 
except those whose BT-transformation is $t_A$, 
and therefore we can find another witness (if it exists) $t_B$.

We continue to use this method, finding couples of witnesses, until there is no rule in the set $\R$ anymore, 
or until we are not able to find a new couple of witnesses anymore: in that latter case, 
we remove the remaining $R_i$ rules because they are useless.

So, now we use the final version of $\A'$ obtained in order to find a term leading to a final state, 
and since we have a couple of witnesses for each rule formerly in the set $\R$, 
we can BT-transform it into a term accepted by $\A^\new$ (hence by $\A$). 
If such a term does not exist, the language recognized by $\A^\new$ 
(i.e. the language recognized by $\A$) is empty.\cqfd
\end{proof}

\section{Conclusion}
Having a tree memory structure instead of a stack is sometimes more relevant
(even when the input functions symbols are only of arities 1 and 0).
We have shown how to extend the visibly pushdown languages
to such memory structures, keeping determinization and closure properties of VPL. 
Our second contribution is then to extend this automaton model,
constraining the transition rules with some regular conditions on memory contents.
The structural equality and disequality
tests appear to a be a good class of constraints since we have then both
decidability of emptiness and Boolean closure properties. 
Moreover, they can be combined (while keeping decidability and closure results)
with equality and disequality tests a la~\cite{BogaertTison92}, 
operating on brothers subterms of the term read.

Several further studies can be done on the automata of this paper.
For instance, the problem of the closure of the corresponding tree languages
under certain classes of term rewriting systems is particularly interesting, 
as it can be applied to the verification of infinite state systems
with \emph{regular model checking} techniques.
It could be interesting as well to study how the definition of VTAM can 
be extended to deal with unranked trees, with the perspective
of applications to problems related to semi-structured documents processing. 


\subsection*{Acknowledgments}
The authors wish to thank Pierre Réty for having noted some mistakes in the examples in the extended abstract, 
and for having sent us a basis of comparison of VTAM with (top down)  Visibly Pushdown Tree Automata, 
and Jean Goubault-Larrecq for his suggestion to refer to 
$\mathcal{H}_3$~\cite{Nielson02sas}
in the proof of Theorem~\ref{th:vtam-emptiness},
and 
the reviewers for their useful and numerous remarks and suggestions.


\begin{thebibliography}{10}

\bibitem{AlurChaudhuriMadhusudan06VPTL}
R.~Alur, S.~Chaudhuri, and P.~Madhusudan.
\newblock Visibly pushdown tree languages.
\newblock Available on: \url{http://www.cis.upenn.edu/~swarat/pubs/vptl.ps},
  2006.

\bibitem{AlurMadhusudan04VPL}
R.~Alur and P.~Madhusudan.
\newblock Visibly pushdown languages.
\newblock In L.~Babai, editor, {\em Proceedings of the 36th Annual ACM
  Symposium on Theory of Computing (STOC 2004)}, pages 202--211. ACM, 2004.

\bibitem{bachmair01hb}
L.~Bachmair and H.~Ganzinger.
\newblock Resolution theorem proving.
\newblock In A.~Robinson and A.~Voronkov, editors, {\em Handbook of Automated
  Reasoning}, chapter~2. North Holland, 2001.

\bibitem{BogaertTison92}
B.~Bogaert and S.~Tison.
\newblock {Equality and Disequality Constraints on Direct Subterms in Tree
  Automata}.
\newblock In {\em 9th Symp. on Theoretical Aspects of Computer Science, STACS},
  volume 577 of {\em LNCS}, pages 161--171. Springer, 1992.

\bibitem{ChabinRety07}
J.~Chabin and P.~R{\'e}ty.
\newblock Visibly pushdown languages and term rewriting.
\newblock In {\em Proc. 6th International Symposium Frontiers of Combining
  Systems (FroCoS)}, volume 4720 of {\em Lecture Notes in Computer Science},
  pages 252--266. Springer, 2007.

\bibitem{charatonik97lics}
W.~Charatonik and A.~Podelski.
\newblock Set constraints with intersection.
\newblock In {\em Proc. IEEE Symposium on Logic in Computer Science}, Varsaw,
  1997.

\bibitem{ComonCortier05tcs}
H.~Comon and V.~Cortier.
\newblock Tree automata with one memory, set constraints and cryptographic
  protocols.
\newblock {\em Theoretical Computer Science}, 331(1):143--214, Feb. 2005.

\bibitem{tata}
H.~Comon, M.~Dauchet, R.~Gilleron, F.~Jacquemard, D.~Lugiez, S.~Tison, and
  M.~Tommasi.
\newblock {\em {Tree Automata Techniques and Applications}}.
\newblock \url{http://www.grappa.univ-lille3.fr/tata}, 1997.

\bibitem{ComonJacquemard03}
H.~Comon and F.~Jacquemard.
\newblock Ground reducibility is {EXPTIME}-complete.
\newblock {\em Information and Computation}, 187(1):123--153, 2003.

\bibitem{CoquideDauchetGilleronVagvolgyi94}
J.-L. Coquid{\'e}, M.~Dauchet, R.~Gilleron, and S.~V{\'a}gv{\"o}lgyi.
\newblock Bottom-up tree pushdown automata: classification and connection with
  rewrite systems.
\newblock {\em Theoretical Computer Science}, 127(1):69--98, 1994.

\bibitem{DershowitzJouannaud90}
N.~Dershowitz and J.-P. Jouannaud.
\newblock {\em Rewrite systems}, chapter Handbook of Theoretical Computer
  Science, Volume B, pages 243--320.
\newblock Elsevier, 1990.

\bibitem{FruhwirthShapiroVardiYardeni91}
T.~Fr\"uhwirth, E.~Shapiro, M.~Vardi, and E.~Yardeni.
\newblock Logic programs as types for logic programs.
\newblock In {\em Proc. of the 6th IEEE Symposium on Logic in Computer
  Science}, pages 300--309, 1991.

\bibitem{goubault06tata}
J.~Goubault-Larrecq.
\newblock R\'esolution ordonn\'ee avec s\'election et classes d\'ecidables en
  logique du premier ordre.
\newblock Lecture Notes, 2006.
\newblock avalaible at
  \url{http://www.lsv.ens-cachan.fr/~goubault/SOresol.pdf}.

\bibitem{Guessarian83}
I.~Guessarian.
\newblock Pushdown tree automata.
\newblock {\em Theory of Computing Systems}, 16(1):237--263, 1983.

\bibitem{Habermehl06tacas}
P.~Habermehl, R.~Iosif, and T.~Vojnar.
\newblock Automata-based verification of programs with tree updates.
\newblock In {\em Proc. 12th Intern. Conf. on Tools and Algorithms for the
  Construction and Analysis of Systems (TACAS'06)}, volume 3920 of {\em LNCS},
  April 2006.

\bibitem{JensenMetayerThorn99}
T.~Jensen, D.~L. M{\'e}tayer, and T.~Thorn.
\newblock Verification of control flow based security policies.
\newblock In {\em Proceedings of the IEEE Symposium on Research in Security and
  Privacy}, pages 89--103. IEEE Computer Society Press, 1999.

\bibitem{Kapur97}
D.~Kapur.
\newblock {\em Essays in Honor of Larry Wos}, chapter Constructors can be
  Partial Too.
\newblock MIT Press, 1997.

\bibitem{Misra94powerlist}
J.~Misra.
\newblock Powerlist: {A} structure for parallel recursion.
\newblock {\em ACM Transactions on Programming Languages and Systems},
  16(6):1737--1767, November 1994.

\bibitem{Nielson02sas}
F.~Nielson, H.~R. Nielson, and H.~Seidl.
\newblock Normalizable horn clauses, strongly recognizable relations and spi.
\newblock In {\em Proc. 9th Static Analysis Symposium (SAS)}, volume 2477 of
  {\em LNCS}, pages 20--35, 2002.

\bibitem{nieuwenhuis01hb}
R.~Nieuwenhuis and A.~Rubio.
\newblock Paramodulation-based theorem proving.
\newblock In A.~Robinson and A.~Voronkov, editors, {\em Handbook of Automated
  Reasoning}, chapter~7. North Holland, 2001.

\bibitem{SchimpfGallier85}
K.~M. Schimpf and J.~Gallier.
\newblock Tree pushdown automata.
\newblock {\em Journal of Computer and System Sciences}, 30(1):25--40, 1985.

\end{thebibliography}



\newpage
\appendix

\section*{Appendix: Two-way  tree automata with structural equality
constraints are as expressive as standard tree automata.}
\label{app:2ways}

In this section, we complete the proof of Lemma~\ref{lem:2ways}.
We show actually a more general result: we consider two-way alternating 
tree automata with some regular constraints  and show that the language they
recognize is also accepted by a standard tree automaton. This
generalizes the proof for two-way alternating tree automata 
(see e.g.~\cite{tata} chapter 7) and the
proof for two-way automata with equality tests~\cite{ComonCortier05tcs},
which itself relies on a transformation from two-way automata to one-way
automata \cite{charatonik97lics}.

Two-way automata are, as usual, automata that can move up and down
and alternation consists (as usual) in spawning to copies of the tree
in different states, requiring acceptance of both copies. In the logical
formalism, alternation simply corresponds to clauses 
$q_1(x),q_2(x) \rightarrow q(x)$,
requiring to accept $x$ both in state $q_1$ and in state $q_2$ if one wants
to accept $x$ in state $q$.

For simplicity, we assume that all function symbols have arity 0 or 2.
Lexical conventions:
\begin{enumerate}[$\bullet$]
\item  $f,g,h,...$ are ranging over symbols of arity 2. Unless explicitly
stated they may denote identical symbols.
\item $a,b,c...$ range over  constants
\item $x, x_1,\ldots, x_i,\ldots, y, \ldots, y_i, z,\ldots, z_i,\ldots$ are
(universally quantified) first-order variables,
\item $S,S_1,S_2,\ldots,S_i,\ldots$ range over states symbols for a fixed
given tree automaton
\item $Q,Q_1,Q_2,\ldots, $ range over states symbols of the tree automaton 
with memory
\item $R,R_1,R_2,\ldots, $ range over state symbols of the binary recognizable
relations.
\end{enumerate}

We assume that $R_i$ are recognizable relations defined by clauses of the
form:
\[
\begin{array}{lrcl}
(A) & & \Rightarrow & \sel{R(a,b)}\\
(B) & S_1(x), S_2(y) & \Rightarrow & \sel{R(f(x,y),a)}\\
(C) & S_1(x), S_2(y) & \Rightarrow & \sel{R(a,f(x,y))}\\
(D) & R_1(x_1,x_2), R_2(y_1,y_2) & \Rightarrow &\sel{ R_3(f(x_1,y_1), g(x_2,y_2))}\\
(E) & S_1(x), S_2(y) & \Rightarrow & \sel{S(f(x,y))}\\
(F) & & \Rightarrow & \sel{S(a)}
\end{array}
\]
We assume wlog that there is a state $S_\top$ in which all trees are accepted (a ``trash state'').

Moreover, we will need in what follows an additional property of the $R_i$'s:
\[ \forall i,j, \exists k, l,  \; R_i(x,y) \wedge R_j(y,z) 
\;\logeq\; R_k(x,y) \wedge R_l(x,z) \]

This property is satisfied by the structural equivalence, for which there
is only one index $i$: $R_i = \equiv$ and we have indeed
\[ x\equiv y \wedge y \equiv z \;\logeq\; x \equiv y \wedge x \equiv z\]
It is also satisfied by the universal binary relation and by the equality
relation. That is why this generalizes corresponding results of
\cite{tata,ComonCortier05tcs}.

Our automata are defined by a finite set of clauses of the form:

\[
\begin{array}{lrcl}
(1) & \sel{Q_1(y_1)}, \sel{Q_2(y_2)}, R(y_1,y_2) & \Rightarrow & Q_3(y_1)\\
(2) & Q_1(y_1), Q_2(y_2) & \Rightarrow & \sel{Q_3(f(y_1,y_2))}\\
(2b) & & \Rightarrow & \sel{Q_1(a)}\\
(3) & \sel{Q_1(f(y_1,y_2))}, Q_2(y_3) & \Rightarrow & Q_3(y_1)\\
(4) & \sel{Q_1(f(y_1,y_2))}, Q_2(y_3) & \Rightarrow & Q_3(y_2)
\end{array}
\]

These clauses have a least Herbrand model. We write $\sem{Q}$ the interpretation of $Q$ in this model. This is the language recognized by the automaton in
state $Q$. 

The goal is to prove that, for every $Q$,
 $\sem{Q}$ is recognized by a finite tree automaton
We use a selection strategy, with splitting and complete the rules
(1)-(4) above. We show that the completion terminates and that we get out of
it  a tree automaton which accepts exactly the memory contents. 
Splitting will introduce nullary predicate symbols (propositional variables).

We consider the following selection strategy.
Let $E_1$ be the set of literals which contain at least one function symbol
and $E_2$ be the set of negative literals


\begin{enumerate}[(1)]
\item If the clause contains a negative literal $\neg R(u,v)$ or a negative
literal $\neg S(u)$ where
either $u,v$ is not a variable, then select such literals only. This
case is ruled out in what follows
\item If the clause contains at least one negated propositional variable,
select the negated propositional variables only. This case is ruled out
in what follows
\item If $E_1\cap E_2\neq \emptyset$, then select $E_1\cap E_2$
\item If $E_1\neq \emptyset$ and $E_1\cap E_2 =\emptyset$, then select
$E_1$
\item If $E_1= \emptyset$ and $E_2\neq \emptyset$, then select the negative 
literals $\neg R(x,y)$ and $\neg S(x)$ if any, otherwise select $E_2$
\item Otherwise, select the only literal of the clause
\end{enumerate}
In what follows (and precedes), selected literals are underlined.

We introduce the procedure by starting to run the completion with the
selection strategy, before showing the general form of the clauses
we get.

First, clauses of the form (3), (4) are replaced (using splitting) with
clauses of the form
\[ 
\begin{array}{lrcl}
(3) & Q_1(f(y_1,y_2)), \sel{\NE{Q_2}} & \Rightarrow & Q_3(y_1)\\
(4) & Q_1(f(y_1,y_2)), \sel{\NE{Q_2}} & \Rightarrow & Q_3(y_2) \\
(s_1)    & \sel{Q_2(x)} & \Rightarrow & \NE{Q_2}
\end{array}
\]

Overlapping $(s_1)$ and (2, 2b) may yield clauses of the form
\[ 
\begin{array}{lrcl}
(s_2) & \sel{\NE{Q_1}}, \sel{\NE{Q_2}} & \Rightarrow & \NE{Q_3}\\
 (s_3) & & \Rightarrow & \sel{\NE{Q}}
\end{array}
\]
together with new clauses of the form $(s_1)$. Eventually, we may reach,
using $(s_3)$ and (3-4) clauses:

\[
\begin{array}{lrcl}
(3b) & \sel{Q_1(f(y_1,y_2))}  & \Rightarrow & Q_3(y_1)\\
(4b) & \sel{Q_1(f(y_1,y_2))} & \Rightarrow & Q_3(y_2) \\
\end{array}
\]

(1) + (2) yields clauses of the form

\[ 
\begin{array}{lrcl}
(5.1) & Q_1(y_1), Q_2(y_2),  \sel{Q_3(g(y_3,y_4))}, R_1(y_1,y_3), R_2(y_2,y_4) & \Rightarrow & Q_4(f(y_1,y_2))\\
(5.2) & Q_1(y_1), Q_2(y_2),  \sel{Q_3(a)}, S_1(y_1), S_2(y_2) & \Rightarrow & Q_4(f(y_1,y_2))\\
(5.3) & \sel{Q_1(a)} & \Rightarrow & Q_2(b)\\
(5.4) & S_1(y_1),S_2(y_2), \sel{Q_1(f(y_1,y_2))} & \Rightarrow & Q_2(a) \\
\end{array}
\]



(2) +(3b) and (2) + (4b) yield clauses of the form (after splitting):

\[ 
\begin{array}{lrcl}
(6) & \sel{\NE{Q_3}}, Q_1(y_1)
& \Rightarrow & 
Q_2(y_1)
\end{array}
\]
and eventually
\[ 
\begin{array}{lrcl}
(6b) & \sel{Q_1(y_1)}
& \Rightarrow & 
Q_2(y_1)
\end{array}
\]

(5.1) + (2)  yields

\[ 
\begin{array}{lrcl}
(7.1) & Q_1(y_1), Q_2(y_2),  Q_3(y_3), Q_4(y_4), R_1(y_1,y_3), R_2(y_2,y_4) & \Rightarrow & \sel{Q_5(f(y_1,y_2))}\\
\end{array}
\]

We split (7.1) : we introduce new predicate symbols $Q_i^{R_j}$ defined by
\[ Q_i(y), R_j(x,y) \Rightarrow Q_i^{R_j}(x) \]
Then clauses (7.1) becomes:

\[ 
\begin{array}{lrcl}
(7.1) & Q_1(y_1), Q_2(y_2),  Q_3^{R_1}(y_1), Q_4^{R_2}(y_2) & \Rightarrow & \sel{Q_5(f(y_1,y_2))}\\
\end{array}
\]

(5.2) + (2b) yields clauses of the form
\[
\begin{array}{lrcl}
(7.2) & Q_1(y_1), Q_2(y_2),  S_1(y_1), S_2(y_2) & \Rightarrow & \sel{Q_3(f(y_1,y_2))}\\
\end{array} 
\]

(6b) + (2) yields new clauses of the form (2).
%
%
(7.1) + (5.1) yields clauses of the form:

\[
\begin{array}{ll}
(8.1) & Q_1(y_1), Q_2(y_2), Q_3^{R_3}(y_1), Q_4^{R_4}(y_2), Q_5(y_3), Q_6(y_4), R_1(y_3,y_1), R_2(y_4,y_2) \\
 & \quad \Rightarrow \sel{Q_7(f(y_3,y_4))}
\end{array} 
\]

At this point, we use the property of $R$ and split the clause:
\[ \exists y_1. Q_1(y_1) \wedge Q_3^{R_3}(y_1) \wedge R_1(y_3,y_1)
\;\logeq\; Q_1^{R_4}(y_1) \wedge Q_3^{R_5}(y_1) \]

Hence clauses (9.1) can be rewritten into clauses of the form:

\[
\begin{array}{lrcl}
(8.1) & Q_1^{R_1}(y_1), Q_3^{R_3}(y_1), Q_5(y_1), Q_2^{R_2}(y_2), Q_4^{R_4}(y_2), Q_6(y_2) & \Rightarrow & \sel{Q_7(f(y_1,y_2))}
\end{array} 
\]

Finally, if we let $\Q$ be the set of predicate symbols consisting of 
\begin{enumerate}[$\bullet$]
\item Symbols $S_i$ 
\item Symbols $Q_i$
\item Symbols $Q_i^{R_j}$ 
\end{enumerate}
For every subset $\mathcal{S}$ of $\Q$, we introduce a propositional variable
$\NE{\mathcal{S}}$. Clauses are split, introducing new propositional variables
(or predicate symbols $Q_i^{R_j}$) in such a way that in all clauses
except split clauses, the variables occurring on the left, also occur on 
the right of the clause. And, in split clauses, there is only one variable
occurring on the left and not on the right.

We let $\mathcal{C}$ be the set of clauses obtained by repeated applications
of resolution with splitting, with the above selection strategy 
(a priori $\mathcal{C}$ could be infinite).
We claim that all generated clauses are of one of the following forms 
(Where the $P_i$'s and the $P'_i$'s belong to $\Q$, $Q$'s states might
actually be $Q_i^{R_j}$)

\paragraph{1. Pop clauses} (the original clauses, which are not subsumed by the new clauses):

\[
\begin{array}{lrcl}
(3) & Q_1(f(y_1,y_2)), \sel{\NE{Q_2}} & \Rightarrow & Q_3(y_1)\\
(4) & Q_1(f(y_1,y_2)), \sel{\NE{Q_2}} & \Rightarrow & Q_3(y_2)\\
(3b) & \sel{Q_1(f(y_1,y_2))} & \Rightarrow & Q_2(y_1)\\
(4b) & \sel{Q_1(f(y_1,y_2))} & \Rightarrow & Q_2(y_2)\\
\end{array}
\]

Note that, clause (1) is a particular case of the alternating clauses below,
since it can be written
\[ Q_1(y_1), Q_2^{R}(y_1) \Rightarrow Q_3(y_1) \] 

\paragraph{2. Push clauses}
\[
\begin{array}{lrcl}
(\textsf{P}_1) & 
 P_1(x),\ldots,P_{n}(x),
P'_1(y),\ldots,P'_{m}(y)  & \Rightarrow & \sel{Q(f(x,y))}\\
(\textsf{P}_2) &
 & \Rightarrow & \sel{P(a)}\\
(\textsf{P}_3) & 
\sel{\NE{\mathcal{S}}}, P_1(x),\ldots,P_{n}(x),
P'_1(y),\ldots,P'_{m}(y)  & \Rightarrow & Q(f(x,y))\\
(\textsf{P}_4) &
\sel{\NE{\mathcal{S}}} & \Rightarrow & Q(a)
\end{array}
\]

\paragraph{3. Intermediate clauses}
\[
\begin{array}{lrcl}
%
(\textsf{I}_1) & 
 P_1(x),\ldots,P_{n}(x),
P'_1(y),\ldots,P'_{m}(y),
 \sel{P''_1(f(x,y))}, \ldots, \sel{P''_{k}(f(x,y))}  & \Rightarrow & Q(f(x,y))\\
(\textsf{I}_2) & 
\sel{P_1(a)},\ldots, \sel{P_{n}(a)}  & \Rightarrow & Q(a)\\
(\textsf{I}_3) & 
 S_1(x_1), S_2(x_2), \sel{Q_1(a)} &
\Rightarrow & Q_2(g(x_1,x_2))\\
(\textsf{I}_4) & 
  \sel{Q_1(a)} &
\Rightarrow & Q_2(b)\\
\end{array}
\]

\paragraph{4. Alternating clauses}
\[
\begin{array}{lrcl}
(\textsf{A}_1) & \sel{\NE{\mathcal{S}}},P_1(x),\ldots,
P_{n}(x) & \Rightarrow & Q(x) \\
(\textsf{A}_2) & \sel{P_1(x)},\ldots,
\sel{P_{n}(x)} & \Rightarrow & Q(x) 
\end{array}
\]

In addition, we have clauses obtained by splitting:

\paragraph{5. Split clauses}
\[
\begin{array}{lrcl}
(\textsf{S}_1)\!\!\!\!
 & \sel{R_j(x,y)}, Q_i(y) & \!\!\Rightarrow\!\! & Q_i^{R_j}(x) \\
(\textsf{S}_1b)\!\!\!\!
 & \sel{R_j(y,x)}, Q_i(y) & \!\!\Rightarrow\!\! & Q_i^{-R_j}(x) \\
(\textsf{S}_2)\!\!\!\!
 & R_1(x_1,y_1), R_2(x_2,y_2), \sel{Q_i(f(y_1,y_2))} &
\!\!\Rightarrow\!\! & Q_i^{\pm R_j}(g(x_1,x_2))\\
(\textsf{S}_3)\!\!\!\!
 & S_1(x), S_2(y), \sel{Q_i(f(x,y))} & \!\!\Rightarrow\!\! & Q_i^{\pm R_j}(a)\!\!\\
(\textsf{S}_4)\!\!\!\!
 & \sel{P_1(x)},\ldots,\sel{P_n(x)} & \!\!\Rightarrow\!\! & \NE{\{P_1,\ldots,P_n\}}\\
(\textsf{S}_5)\!\!\!\!
 & P_1(x),\ldots,P_n(x),P'_1(y),\ldots,P'_m(y),
\sel{P''_1(f(x,y))}, \ldots, \sel{P''_k(f(x,y))} & \!\!\Rightarrow\!\! & \NE{\mathcal{S}}\\
\end{array}
\]

\paragraph{6. Propositional clauses}
\[
\begin{array}{lrcl}
(\textsf{E}_1) & \sel{\NE{\mathcal{S}_1}},\ldots, \sel{\NE{\mathcal{S}_n}} & \Rightarrow & \NE{\mathcal{S}}\\
(\textsf{E}_2) & & \Rightarrow & \sel{\NE{\mathcal{S}}}\\
(\textsf{E}_3) & \sel{P_1(a)}, \ldots \sel{P_n(a)} & \Rightarrow & \NE{\mathcal{S}}
\end{array}
\]

Every resolution step using the selection strategy of two of the above clauses
 yield a clause in the above set
\begin{enumerate}[\ ]
\item {\hskip-8 pt$\POP$+$\PUSH$:} yields an alternating clause $(\textsf{A}_1)$ and a split
clause $(\textsf{S}_4)$.
\item {\hskip-8 pt$\INT$ + $\PUSH$:} yields a Push clause or an intermediate clause
\item {\hskip-8 pt$\mathsf{alternating}$ + $\PUSH$:}  yields an intermediate clause 
$(\textsf{I}_1)$ or
$(\textsf{I}_2)$.
\item {\hskip-8 pt$\mathsf{split}$ + $R$:} yields a split clause $(\textsf{S})_2$ or 
$(\textsf{S}_3)$
or an intermediate clause $(\textsf{I}_3)$ or $(\textsf{I}_4)$.
\item {\hskip-8 pt$(\mathsf{S}_2)$ + $\PUSH$:} yields clauses $(\textsf{S}_1)$ and push clauses. Note that here, we use the property of the relation $R$ to split clauses,
which may involve predicates $Q_i^{R_j}$.
\item {\hskip-8 pt$(\mathsf{S}_3)$+ $\PUSH$:} yields push clause and split clauses 
$(\textsf{S}_4)$.
\item {\hskip-8 pt$(\mathsf{S}_4)$+ $\PUSH$:} yields  split clauses $(\textsf{S}_5)$ or
propositional clause $(\textsf{E}_3)$.
\item {\hskip-8 pt$(\mathsf{S}_5)$+ $\PUSH$:} yields  split clauses $(\textsf{S}_5)$ or
propositional clause $(\textsf{E}_1)$.
\end{enumerate}

It follows that all clauses of $\mathcal{C}$ are of the above form. Since there
are only finitely many such clauses, $\mathcal{C}$ is finite and computed
in finite (exponential) time.

Now, we let $\A$ be the alternating tree automaton defined by
clauses $(\textsf{P}_1)$ and $(\textsf{P}_2)$ (and automata clauses defining
the $S$ states). Let, for any state $Q$, $\sem{Q}_{\A}$ be the language
accepted in state $Q$ by $\A$. We claim that $\sem{Q}=\sem{\A}$.

To prove this, we first show (the proof is omitted here) that 
$\NE{\{P_1,\ldots,P_n\}}$ is in $\mathcal{C}$ 
iff $\sem{P_1}_{\A} \cap \ldots \cap \sem{P_n}_{\A}\neq \emptyset$.

Then observe that $\sem{Q}$ is also the interpretation of $Q$ in the least
Herbrand model of $\mathcal{C}$: indeed, all computations yielding $\mathcal{C}$ are
correct. 
Since $\sem{Q}_{\A} \subseteq \sem{Q}$ is trivial, we only have to prove
the converse inclusion. For every $t\in \sem{Q}$ there is a proof of $Q(t)$
using the clauses in $\mathcal{C}$. 

Assume, by contradiction, that there is a term $t$ and a predicate
symbol $Q$ such that all proofs of $Q(t)$ using the clauses
in $\mathcal{C}$ involve at least a clause, which is not an automaton clause.
Then, considering an appropriate sub-proof, there is a term $u$ and
a predicate symbol $P$ such that all proofs of $P(u)$ involve at least
one non-automaton clause and there is a proof of $P(u)$ which uses
exactly one non-automaton clause, at the last step of the proof.


We investigate all possible cases for the last clause used in the proof of $P(u)$
 and derive a contradiction in each case.






\begin{enumerate}[\ ]

\item {\hskip-8 pt\bf Clause $\textsf{I}_1$:} The last step of the proof is
\[
\inference{P_1(u_1),\ldots, P_n(u_1), P'_1(u_2),\ldots,P'_m(u_2),
P''_1(f(u_1,u_2)), \ldots, P''_k(f(u_1,u_2))}{P(f(u_1,u_2))}
\]
and we assume $u=f(u_1,u_2)$. Assume also that, among the proofs we consider,
$k$ is minimal. (If $k=0$ then we have a push clause, which is supposed
not to be the case).

By hypothesis,  for all $i$, 
$u_1\in \sem{P_i}_{\A}$, $u_2\in \sem{P'_i}_{\A}$ and
$f(u_1,u_2)\in \sem{P''_i}_{\A}$. In particular, if we
consider the last clause used in the proof of $P''_k(u)$:

\[Q_1(x),\ldots, Q_r(x), Q'_1(y),\ldots,Q'_s(y) \Rightarrow P''_k(f(x,y)) \]
belongs to $\mathcal{C}$. Then, overlapping this clause with the above clause
$\textsf{I}_1$, the following clause belongs also to $\mathcal{C}$:
\[ \begin{array}{l}
P_1(x),\ldots,P_n(x),Q_1(x),\ldots,Q_r(x),\\
P'_1(y),\ldots,P'_m(y), Q'_1(y),
\ldots,Q'_s(y), P''_1(f(x,y)),\ldots, P''_{k-1}(f(x,y)) \Rightarrow P(f(x,y))
\end{array}\]
 and therefore we have another proof of $P(u)$:
\[ \inference{\begin{array}{l}
P_1(u_1),\ldots, P_n(u_1), Q_1(u_1),\ldots,Q_r(u_1) \\
P'_1(u_2),\ldots,P'_m(u_2), Q'_1(u_2),\ldots,Q'_s(u_2),
P''_1(f(u_1,u_2)), \ldots, P''_{k-1}(f(u_1,u_2))\end{array}}{P(f(u_1,u_2))} 
\]

which contradicts the minimality of $k$.

\item {\hskip-8 pt\bf Clause $(\textsf{A}_1)$:} The last step of the proof is
\[ \inference{P_1(u), \ldots, P_n(u)}{P(u)} \]
By hypothesis, the proofs of $P_i(u)$ only use automata clauses:
$\forall i. u\in \sem{P_i}_{\A}$. Le the push rule
\[ Q_1(x), \ldots, Q_m(x), Q'_1(y),\ldots,Q'_p(y) \Rightarrow P_n(f(x,y)) \]
be the last clause used in the proof of $P(u)$.
Overlapping this clause and the clause $\textsf{A}_1$ above, there is 
another clause in $\mathcal{C}$ yielding a proof of $P(u)$:
\[ Q_1(x), \ldots, Q_m(x), Q'_1(y),\ldots,Q'_p(y), P_1(f(x,y)), \ldots, P_{n-1}(f(x,y)) \Rightarrow P(f(x,y)) \]

And we are back to the case of $\textsf{I}_1$.

\item {\hskip-8 pt\bf Clause (3b):}
\[
\inference{Q_1(f(u,t))}{P(u)}
\]
By hypothesis 
 $f(t,u)\in \sem{Q_1}_{\A}$. Hence there is a push clause
\[ P_1(x), \ldots, P_n(x), P'_1(y),\ldots,P'_m(y) \Rightarrow Q_1(f(x,y)) \]
such that $t\in \sem{P_1}_{\A} \cap \ldots \cap \sem{P_n}_{\A}$
and $u\in \sem{P'_1}_{\A} \cap \ldots \cap \sem{P'_m}_{\A}$. 
By resolution on the clause (3b), there is also in $\mathcal{C}$ a clause
\[ P_1(x), \ldots, P_n(x), \NE{\{P'_1,\ldots,P'_m\}} \Rightarrow Q(x) \]
However, since 
$\sem{P'_1}_{\A}\cap \ldots \cap \sem{P'_m}_{\A}\neq \emptyset$, $\NE{\{P'_1,\ldots,p'_m\}}$ 
is also in $\mathcal{C}$ and, by resolution
again
\[ P_1(x), \ldots, P_n(x)\Rightarrow Q(x) \]
 is a clause of $\mathcal{C}$.

Then we are back to the case of $\textsf{A}_1$.

\item {\hskip-8 pt\bf Clause (3):}   The last step of the proof is
\[
\inference{\juxtapose{Q_1(f(u,t))}{\NE{Q_2}}}{P(u)}
\]
Since $\NE{Q_2}\in \mathcal{C}$ in this case, by saturation of $\mathcal{C}$, there
is a clause $Q_1(x,y) \Rightarrow Q(x)$ in $\mathcal{C}$, and we are back to the
case of $(3b)$. 



\item {\hskip-8 pt\bf Other cases:} they are quite similar to the previous
ones. Let us only  consider the case of clause $(\textsf{S}_2)$, which is
slightly more complicated.
\[ \inference{R_1(u_1,v_1)\;  R_2(u_2,v_2)\; Q_i(f(v_1,v_2))}{Q_i^{R_j}(g(u_1,u_2))}\]
Assume moreover that $u=g(u_1,u_2)$ is a minimal size term such that, for
some $Q_i, R_j$,
$Q_i^{R_j}(u)$ is provable using as a last step
an inference $\textsf{S}_2$, and  is
not provable by automata clauses only,

As before, we consider the overlap between $\textsf{S}_2$ and a push clause.
We get
\[ R_1(x_1,y_1), R_2(x_2,y_2), P_1(y_1), \ldots, P_n(y_1), P'_1(y_2),\ldots,
P'_m(y_2) \Rightarrow Q_i^{R_j}(g(x_1,x_2)) \] 
Hence, the following clauses belong to $\mathcal{C}$ (when $P_i$, $P'_i$
are not themselves
predicates $Q^R$; otherwise, we have to use the property on $R$ relations
and split in another way, using the $S_\top$ predicate, as shown later):
\[ \begin{array}{rcl}
\sel{R_1(x_1,y_1)}, P_i(y_1) & \Rightarrow & P_i^{R_1}(x_1)\\
\sel{R_2(x_2,y_2)}, P'_i(y_2) & \Rightarrow & {P'_i}^{R_2}(x_2)\\
P_1^{R_1}(x_1),\ldots, P_n^{R_1}(x_1), {P'_1}^{R_2}(x_2),\ldots,{P'_m}^{R_2}(x_2)
& \Rightarrow & \sel{Q_i^{R_j}(g(x_1,x_2))} 
\end{array}
\] 
and we have the following proof of $g(u_1,u_2)$:

\[
\prooftree
\prooftree
R_1(u_1,v_1)\;
P_1(v_1)
\justifies P_1^{R_1}(u_1)
\endprooftree 
\cdots
\prooftree
R_1(u_1,v_n)\;
P_n(v_1)
\justifies P_n^{R_1}(u_1)
\endprooftree
\prooftree
R_2(u_2,w_1)\;
P'_2(w_1)
\justifies {P'_1}^{R_2}(u_2)
\endprooftree
\cdots 
\prooftree
R_2(u_2,w_m)\;
P'_m(w_m)
\justifies {P'_m}^{R_2}(u_2)
\endprooftree
\justifies
Q_i^{R_j}(g(u_1,u_2))
\endprooftree
\]

Now, by overlapping again $R_1(x_1,y_1)$ and $R_2(x_2,y_2)$ with their
defining clause, we compute ``shortcut clauses'' belonging to $\mathcal{C}$ and
get another proof (for instance assuming $v_1= f(v_{11},v_{12})$ and
$u_1 = h(u_{11}, u_{12})$):

\[
\prooftree
\prooftree
R_{11}(u_{11},v_{11})\;
R_{12}(u_{12},v_{12})\;
P_1(f(v_{11},v_{12}))
\justifies P_1^{R_1}(u_1)
\endprooftree 
\cdots
\prooftree
R_2(u_2,w_1)\;
P'_2(w_1)
\justifies {P'_1}^{R_2}(u_2)
\endprooftree
\cdots 
\prooftree
R_2(u_2,w_m)\;
P'_m(w_m)
\justifies {P'_m}^{R_2}(u_2)
\endprooftree
\justifies
Q_i^{R_j}(g(u_1,u_2))
\endprooftree
\]

By minimality of $u$, $u_1\in \sem{P_1^{R_1}}_{\A}$. Similarly,
for every $i$, $ u_1\in \sem{P_i^{R_1}}_{\A}$. $u_2\in \sem{{P'_i}^{R_2}}_{\A}$ and it follows that $g(u_1,u_2) \in \sem{Q_i^{R_j}}_{\A}$.

Finally, let us consider the case where some $P_i$ is itself a predicate
symbol $Q^R$, in which case we do not have a predicate $(Q^R)^{R_1}$. 
We use then the assumed property of the predicates $R_i$: 
$R_1(x,y) \wedge R(y,z) \logeq R'_1(x,y) \wedge R'(x,z)$, hence
\[ (\exists u, \exists v.  R_1(x,u) \wedge R(u,v) \wedge Q(v))
\logeq (\exists u. R_1(x,u) \wedge S_\top(u)) \wedge (\exists v. R(x,v) 
\wedge Q(v)) \]
Hence we need two split clauses instead of one:
\[ \begin{array}{rcl}
R'_1(x,y) & \Rightarrow & S_\top^{R'_1}(x)\\
R'(x,y), Q(y) & \Rightarrow & Q^{R'}(x)\\
\end{array}
\]
And $R_1(x_1,y_1), Q^R(y_1)$ is replaced with $S_\top^{R'_1}(x_1), Q^{R'}(x_1)$.
Note that such a transformation is not necessary when there is a single
transitive binary relation, as in our application: then $R(x,y) \wedge Q^R(y)$
is simply replaced with $Q^R(x)$. 

\end{enumerate}

To sum up: if there is a proof of $P(u)$ using clauses of $\mathcal{C}$, then,
by saturation of the clauses of $\mathcal{C}$ w.r.t. overlaps with push clauses,
we can rewrite the proof into a proof using push clauses only:
$u\in \sem{P}_{\A}$. This proves that $\sem{P}=\sem{P}_{\A}$.

Finally, it is easy (and well-known) to compute a standard bottom-up
automaton accepting the same language as an alternating automaton;
this only requires a subset construction. That is why the language 
accepted by our two-way automata with structural equality constraints
is actually a recognizable language. The overall size of the resulting
automaton (and its computation time) are simply exponential, but we 
know that, already for alternating automata, we cannot do better.
\end{document}